\documentclass[11pt,a4paper]{article}
\usepackage{jheppub}

\usepackage[english]{babel}
\usepackage{graphicx}

\usepackage{amsmath,amsfonts,amssymb}
\usepackage{calrsfs}     
\usepackage{dsfont}      
\usepackage{enumerate}

\newcommand\Lagr{\mathcal{L}}
\newcommand\identity{\mathds{1}}
\renewcommand{\bar}{\overline}
\DeclareMathOperator{\tr}{Tr}

\newcommand{\tev}{\,\, \mathrm{TeV}}

\title{The Little Skyrmion: \\New Dark Matter for Little Higgs Models}

\author[a]{Marc Gillioz,}
\author[a]{Andreas von Manteuffel,}
\author[a,b,c]{Pedro Schwaller,}
\author[a]{Daniel Wyler\,}

\affiliation[a]{Institut f\"ur Theoretische Physik,
	Universit\"at Z\"urich, \\  Winterthurerstrasse 190, CH-8057
	Z\"urich, Switzerland}
\affiliation[b]{HEP Division, Argonne National Laboratory,\\
	9700 Cass Ave, Argonne, IL 60439, USA}
\affiliation[c]{Department of Physics, University of Illinois,\\
	845 W Taylor St, Chicago, IL 60607, USA}

\emailAdd{gillioz@physik.uzh.ch}
\emailAdd{manteuffel@physik.uzh.ch}
\emailAdd{pschwaller@hep.anl.gov}
\emailAdd{wyler@physik.uzh.ch}

\abstract{We study skyrmions in the littlest Higgs model and discuss their possible role as dark matter candidates.
Stable massive skyrmions can exist in the littlest Higgs model also in absence of an exact parity symmetry, since they carry a conserved topological charge due to the non-trivial third homotopy group of the $SU(5)/SO(5)$ coset.
We find a spherically symmetric skyrmion solution in this coset.
The effects of gauge fields on the skyrmion solutions are analyzed and found to lead to 
an upper bound on the skyrmion mass. The relic abundance is in agreement with the observed dark matter density for reasonable parameter choices.}

\arxivnumber{1012.5288}

\subheader{ZU-TH 20/10\\
	ANL-HEP-PR-10-68}

\begin{document}
\maketitle

\flushbottom

\section{Introduction}

Little Higgs models are extensions of the standard model, where the Higgs scalar is a pseudo-Nambu-Goldstone boson of a global symmetry $G$, spontaneously broken at a scale $f \sim 1 \tev$ to a subgroup $H$.
The enlarged global symmetry, together with a suitable embedding of gauge and Yukawa interactions, protects the Higgs mass from large radiative corrections at the one loop level, and provides a natural explanation for the hierarchy between the electroweak scale $v$ and the global symmetry breaking scale $f$.
A simple implementation of this mechanism is given by the ``littlest Higgs'' model of ref. \cite{ArkaniHamed:2002qy} which is based on the coset $G/H = SU(5)/SO(5)$. 
Models with other symmetry breaking patterns include the ``minimal moose'' model \cite{ArkaniHamed:2002qx} based on a $[(SU(3)\times SU(3))/SU(3)]^4$ coset, the antisymmetric model using a $SU(6)/Sp(6)$ coset \cite{Low:2002ws}, the $[SO(5)\times SO(5)/SO(5)]^4$ model of \cite{Chang:2003un}, or the ``bestest little Higgs'' with $SO(6)\times SO(6)/SO(6)$ symmetry \cite{Schmaltz:2010ac}. At a scale $\Lambda \approx 4\pi f$ little Higgs models become strongly coupled and must be supplemented by a UV-completion. The global symmetry breaking patterns can be thought to arise from a set of (techni-) fermions which condense due to gauge interactions becoming strongly coupled at the scale $\Lambda$, similar to the mechanism of chiral symmetry breaking in QCD. 

The coset spaces upon which the little Higgs models are based may have nontrivial homotopy groups, which in turn lead to the existence of solitons --- topologically nontrivial field configurations, also known as topological defects. This was already noted in \cite{Trodden:2004ea} where zero- and one-dimensional topological defects, {\it monopoles} and {\it strings}, are shown to exist in the littlest Higgs model, and in \cite{Hill:2007zv} where the possible presence of {\it skyrmions} in the littlest Higgs model is mentioned. Stable skyrmions \cite{Skyrme:1961vq} can exist provided that the third homotopy group $\pi_3 (G/H)$ of the coset is nontrivial. They represent ``baryons'' formed at the scale $\Lambda$ \cite{Adkins:1983ya}. Among the little Higgs models admitting skyrmion solutions are the littlest Higgs where $\pi_3(SU(5)/SO(5)) = {\mathbb Z}_2$ and the minimal moose model with $\pi_3(SU(3)\times SU(3)/SU(3)) = \mathbb Z$ \cite{Hill:2007zv,Murayama:2009nj}. Other models, for example the ones based on $SU(6)/Sp(6)$,  have a trivial third homotopy group and therefore do not possess stable skyrmion solutions.

Due to their nontrivial global structure, skyrmions carry a topological charge.
Their masses and sizes are stabilised by completing the effective Lagrangian with
a particular higher-derivative operator, the so-called Skyrme term~\cite{Skyrme:1961vq}.
This mechanism prevents the lightest skyrmion state from decaying, thus providing a new possible candidate for dark matter in little Higgs models, without requiring the introduction of a parity symmetry by hand. This is a welcome alternative since it was recently shown that T-parity \cite{Cheng:2004yc} is violated by anomalies \cite{Hill:2007nz,Hill:2007zv} and leads to the decay of the dark matter candidate \cite{Barger:2007df,Freitas:2008mq}. While alternative implementations of a dark matter parity are possible (see e.g. \cite{BirkedalHansen:2003mpa,Krohn:2008ye,Csaki:2008se,Freitas:2009jq,Brown:2010ke}), the topological charge also protects the dark matter candidate from possible symmetry breaking induced at higher scales, and allows for simpler low energy structures of little Higgs models. First steps towards skyrmion dark matter for little Higgs models were taken in \cite{Joseph:2009bq}, where a spherically symmetric skyrmion solution is found. 
Notice also that skyrmions have been shown recently to appear in compact five-dimensional models \cite{Pomarol:2007kr}, where they provide a successful description of QCD baryons \cite{Domenech:2010aq}.

Our paper improves and extends the previous analyses in little Higgs models in several ways. After reviewing the littlest Higgs model and introducing the Skyrme term in section 2, in section 3 we present a spherically symmetric skyrmion solution which is significantly lighter than previous solutions. We then show that gauge interactions further reduce the mass of the skyrmion and preserve its stability on cosmological time scales. 
In section 4 we study the skyrmion self-interactions, estimate its annihilation cross-section, and derive cosmological bounds. In section 5 we discuss shortly the presence and properties of skyrmions in other realisations of the little Higgs model. Finally, section 6 contains our conclusions. The limiting behaviour of the gauged skyrmion for a large Skyrme term is discussed in the appendix.


\section{The model}

\subsection{The littlest Higgs}

We consider the littlest Higgs model of ref.~\cite{ArkaniHamed:2002qy}. The model is based on a global $SU(5)$ symmetry, spontaneously broken down to $SO(5)$ by a vacuum expectation value. The Nambu-Goldstone bosons are therefore described by a $SU(5)/SO(5)$ non-linear sigma model
\begin{equation}
	\Lagr_\Sigma = \frac{f^2}{4} \tr \partial_\mu \Sigma \partial^\mu \Sigma^\dag,
\end{equation}
where $\Sigma$ is a $5 \times 5$ symmetric matrix. Under a global $SU(5)$ transformation, $\Sigma$ transforms as $\Sigma \to V \Sigma V^T$, with $V \in SU(5)$. The vacuum expectation value is taken to be the identity matrix, $\langle \Sigma \rangle = \identity_5$, so that the 10 unbroken generators obey $(T^a)^T = -T^a$, and the 14 broken ones $(X^a)^T = X^a$. The Nambu-Goldstone bosons $\pi^a$ can therefore be parametrised as $\Sigma = (e^{i \pi_a X^a / f}) (e^{i \pi_a X^a / f})^T = e^{2 i \pi_a X^a / f}$.

The global $SU(5)$ symmetry is then explicitly broken by gauging an $\left[ SU(2) \times U(1) \right]^2$ subgroup. The generators of this gauge group are chosen as\footnote{
With respect to the generators as defined in ref.~\cite{ArkaniHamed:2002qy}, our generators are rotated according to the rule $Q_i^{(\alpha)} \to \Omega Q_i^{(\alpha)} \Omega^\dag$ (and similarly for the $Y^{(\alpha)}$, where our
definition differs by an additional overall minus sign).
Here, $\Omega$ is a $SU(5)$ matrix taking the vacuum expectation value $\Sigma_0$ of ref.~\cite{ArkaniHamed:2002qy} to the identity, $\Sigma_0 \to \Omega\Sigma_0\Omega^T = \identity_5$, and is defined as
\begin{equation*}
	\Omega = \frac{1}{\sqrt 2} \left( \begin{array}{ccc}
		 \identity_2 & 0 & \identity_2 \\
		0 & -\sqrt 2 & 0 \\
		-i \identity_2 & 0 & i \identity_2
	\end{array} \right).
\end{equation*}}
\begin{eqnarray}
	Q_i^{(1)} = \frac{1}{4} \left( \begin{array}{ccc}
		\sigma_i & 0 & i \sigma_i \\
		0 & 0 & 0 \\
		-i \sigma_i & 0 & \sigma_i
	\end{array} \right), \hspace{1.0cm}
	&\hspace{1.5cm}&
	Q_i^{(2)} = -Q_i^{(1)T},
	\label{equ:Q} \\
	Y^{(1)} = \frac{1}{20} \left( \begin{array}{ccc}
		\identity_2 & 0 & 5 i \identity_2 \\
		0 & -4 & 0 \\
		-5 i \identity_2 & 0 & \identity_2
	\end{array} \right),
	&\hspace{1.5cm}&
	Y^{(2)} = - Y^{(1)T},
	\label{equ:Y}
\end{eqnarray}
and they obey $[ Q_i^{(\alpha)}, Q_j^{(\beta)} ] = i \delta^{\alpha\beta} \epsilon_{ijk} Q_k^{(\alpha)}$, $[ Q_i^{(\alpha)}, Y^{(\beta)} ] = [ Y^{(\alpha)}, Y^{(\beta)} ] = 0$. The commutation relations of the $SU(2)$ and $U(1)$ subgroups is easier to see in the original parametrisation of ref.~\cite{ArkaniHamed:2002qy}, in which the vacuum expectation value $\langle \Sigma \rangle$ is not diagonal. In our case however, it will be more convenient to work in a basis where $\langle \Sigma \rangle = \identity_5$. The Lagrangian is made gauge invariant by promoting the spacetime derivatives to covariant derivatives:
\begin{equation}
	D_\mu \Sigma = \partial_\mu \Sigma - i \left( A_\mu \Sigma + \Sigma A_\mu^T \right),
\end{equation}
where $A_\mu = \sum_{\alpha=1,2} \left( g_\alpha W_\mu^{(\alpha),a} Q_a^{(\alpha)} + g'_\alpha B_\mu^{(\alpha)} Y^{(\alpha)} \right)$.

Only the linear combinations $Q_a = Q_a^{(1)} + Q_a^{(2)}$, $Y = Y^{(1)} + Y^{(2)}$ of the gauge generators are symmetric and thus preserve the vacuum. The orthogonal combinations $\bar Q_a = Q_a^{(1)} - Q_a^{(2)}$, $\bar Y = Y^{(1)} - Y^{(2)}$ do not. The $\left[ SU(2) \times U(1) \right]^2$ gauge group is therefore spontaneously broken down to a diagonal $SU(2) \times U(1)$ subgroup. The latter is identified with the standard model electroweak gauge group.

To simplify the structure of the gauge sector of the model we work in the T-parity symmetric limit \cite{Cheng:2004yc} which is obtained by setting $g_1 = g_2 = \sqrt 2 g$ and $g'_1 = g'_2 = \sqrt 2 g'$. This also allows us to consider values of the breaking scale $f \lesssim 1\tev$. The standard model gauge bosons are identified with the parity even linear combinations
\begin{align}
	 W_\mu^a = \frac{1}{\sqrt 2} \left( W_\mu^{(1),a} + W_\mu^{(2),a} \right)\,, \qquad B_\mu = \frac{1}{\sqrt 2} \left( B_\mu^{(1)} + B_\mu^{(2)} \right)\,.
\end{align}
The parity odd linear combinations 
\begin{align}
	\bar W_\mu^a = \frac{1}{\sqrt 2} \left( W_\mu^{(1),a} - W_\mu^{(2),a} \right)\,, \qquad \bar B_\mu = \frac{1}{\sqrt 2} \left( B_\mu^{(1)} - B_\mu^{(2)} \right),
\end{align}
are responsible for cutting off the quadratically divergent contribution to the Higgs mass in the gauge sector. 
They obtain tree-level masses
\begin{equation}
	m_{\bar W}^2 = 2 g^2 f^2,
	\hspace{2cm}
	m_{\bar B}^2 = \frac{2}{5} g'^2 f^2.
\end{equation}
The 14 Nambu-Goldstone bosons can be parametrised as $\Sigma = e^{2 i \Pi/f}$. They decompose under the electroweak gauge group as $\mathbf{1}_0 \oplus \mathbf{3}_0 \oplus \mathbf{2}_{\pm 1/2} \oplus \mathbf{3}_{\pm 1}$. Explicitly, we have
\begin{equation}
	\Pi = \frac{1}{2\sqrt 2} \left( \begin{array}{ccc}
		-i \phi - \omega - \frac{1}{2\sqrt 5} \eta & -h & -\phi - i \omega \\
		-h^T & \frac{2}{\sqrt 5} \eta & i h^T \\
		-\phi + i \omega & i h & i\phi - \omega - \frac{1}{2\sqrt 5} \eta
	\end{array} \right)
	+ \textrm{c.c.},
\end{equation}
\begin{equation*}
	h = \left( \begin{array}{c}
		h^+ \\ h^0
	\end{array} \right),
	\hspace{0.5cm}
	\phi = \left( \begin{array}{cc}
		\phi^{++} & \frac{1}{\sqrt 2} \phi^+ \\
		\frac{1}{\sqrt 2} \phi^+ & \phi^0
	\end{array} \right),
	\hspace{0.5cm}
	\omega = \left( \begin{array}{cc}
		\frac{1}{2} \omega^0 & \frac{1}{\sqrt 2} \omega^+ \\
		\frac{1}{\sqrt 2} \omega^- & -\frac{1}{2} \omega^0
	\end{array} \right).
\end{equation*}
The real triplet $\omega$ and the singlet $\eta$ are eaten by the Higgsing of the broken $SU(2) \times U(1)$. The complex doublet $h$ is identified with the standard Higgs boson, while the complex triplet $\phi$ is a new field of the model, which receives a large ${\cal O}(gf)$ mass at the one loop level. The degeneracy between the triplet states is lifted after electroweak symmetry breaking by a vacuum expectation value for the Higgs doublet $\langle h \rangle = (0,v/\sqrt 2)^T$, also giving the standard model $W^\pm$ and $Z$ bosons their mass. For more details we refer the reader to \cite{Hubisz:2004ft}.

\subsection{Skyrme term}

We are interested in finding solutions of the classical equations of motion for the field $\Sigma$ with nontrivial topological charge, i.e. solutions which cannot be deformed into the vacuum state $\langle \Sigma \rangle = \identity$ by a series of infinitesimal transformations. 

In order to identify those special field configurations with particles, they need to have a finite energy and size. Notice that we will only consider time-independent field configurations here; propagating solitons can then be obtained from static ones by applying a Lorentz boost. 

The finite energy requirement implies that at large distances from the origin, the field $\Sigma$ must approach the vacuum expectation value: $\Sigma(x\to\infty) = \langle\Sigma\rangle = \identity_5$. For this reason, all points located at spatial infinity can be identified, and the configuration space $\mathbb R^3$ is topologically equivalent to the three sphere $S^3$. The space of solutions to the equations of motion can therefore be split into homotopy classes, characterised by the third homotopy group $\pi_3$. 

For the littlest Higgs coset, we have $\pi_3 (SU(5)/SO(5)) = \mathbb Z_2$ \cite{Bryan:1993hz}, which means that there exist two topologically inequivalent classes of field configurations, characterised by a winding number equal to 0 or 1. Field configurations of winding number zero can be continuously deformed into the identity, but this is not possible for the configurations of winding number one --- they carry a conserved topological charge. Unfortunately, while for $SU(N)$ the winding number can be expressed as a simple integral over spacetime, there is to the best of our knowledge no such universal quantity for $SU(N)/SO(N)$ cosets.

However, one can still ensure by construction that a configuration has a given winding number. If one considers a field $\Phi(x) \in SU(5)$, its winding number (or topological charge) in $SU(5)$ is given by
\begin{equation}
	\eta(\Phi) = -\frac{1}{24 \pi^2} \epsilon_{ijk} \int d^3 x~ \tr (\Phi^\dag \partial_i \Phi )
		( \Phi^\dag \partial_j \Phi) ( \Phi^\dag \partial_k \Phi )
		~ \in \mathbb Z.
	\label{equ:windingnumber}
\end{equation}
The winding number integral is additive: $\eta(\Phi_1 \Phi_2) = \eta(\Phi_1) + \eta(\Phi_2)$. 
Furthermore, for a field $R(x) \in SO(5)$, we have that $\eta(R)$ is an even integer, i.e. $\eta(R) \in 2 \mathbb Z$. Following \cite{Bryan:1993hz} we can then construct a field of winding number one in the coset as follows: given a map $\Phi(x) \in SU(5)$, we write $\Sigma(x) = \Phi(x) \Phi(x)^T$ which defines at each point of space a representative of $SU(5)/SO(5)$. We then have that the quantity
\begin{equation}
    \tilde\eta(\Sigma) = \eta(\Phi) \text{ mod } 2
    \label{equ:winding_coset}
\end{equation}
describes the winding number of the field configuration $\Sigma(x)$. Taking a field configuration with $\eta(\Phi)=1$ therefore ensures a topologically non-trivial field configuration $\Sigma(x)$. A proof of the relation \eqref{equ:winding_coset} is for example given in \cite{Bryan:1993hz} using the exact homotopy sequence. Here we just note that it is consistent with the fact that $\Phi R$ with $R \in SO(5)$ gives the same $\Sigma$ as $\Phi$ by construction, so that $\tilde\eta(\Sigma)$ is uniquely defined.

These field configurations with nontrivial topological charge ensure the existence of skyrmions \cite{Skyrme:1961vq,Adkins:1983ya}.
However, in order to stabilise their energy and size, one need to complete the Lagrangian with terms with higher number of derivatives. The simplest choice is to add the so-called Skyrme term:
\begin{equation}
	\Lagr_{Skyrme} = \frac{1}{32 e^2} \tr \left[ \Sigma^\dag D_\mu \Sigma, \Sigma^\dag D_\nu \Sigma \right]
		\left[ \Sigma^\dag D^\mu \Sigma, \Sigma^\dag D^\nu \Sigma \right].
	\label{equ:Skyrme_term}
\end{equation}
This term is obviously invariant under both $SU(5)$ global transformations and gauge transformations. Moreover, since each of the commutators in \eqref{equ:Skyrme_term} is antisymmetric in its Lorentz indices, the Skyrme term contains at most two time derivatives, which facilitates the quantisation procedure.

The Skyrme term does not modify the mass of the gauge bosons at tree-level, but it induces new couplings between gauge and Nambu-Goldstone bosons. The gauge-boson four vertices are of particular interest since they might contribute at one loop to the electroweak precision measurements. 
In practice, the contributions to the Peskin-Takeuchi $S$ and $T$ parameter \cite{Peskin:1990zt,Peskin:1991sw} are suppressed by the loop factor and by powers of $(v/f)^2$ and are thus negligible as long as $(1/e^2) \lesssim 32$. 

The other important place where the Skyrme term might play a role is in the potential for the scalars. Since the Skyrme term involves four derivatives, these contributions only start at the two loop level. The protection of the Higgs mass through the collective symmetry breaking mechanism is therefore not affected by the addition of the Skyrme term. 

Other terms with four or more derivative might also be included in addition to the Skyrme term~\eqref{equ:Skyrme_term}. In particular, a term with 6 derivatives is often used in the QCD Skyrme models to stabilise the soliton \cite{Adkins:1983nw,Jackson:1985yz}, namely the square of the topological current: $\Lagr_6 \sim \tr B_\mu B^\mu$. However, in the littlest Higgs model, and more generally for a skyrmion living in a $SU(N)/SO(N)$ coset, the topological current built out of the $\Sigma$ fields is vanishing due to the group structure:
\begin{equation}
	B^\mu = \frac{1}{24 \pi^2} \epsilon^{\mu\nu\rho\sigma} \int d^3 x~
		\tr ( \Sigma^\dag \partial_\nu \Sigma ) ( \Sigma^\dag \partial_\rho \Sigma ) ( \Sigma^\dag \partial_\sigma \Sigma ) = 0
\end{equation}
Motivated by QCD, where no other higher derivative terms than the Skyrme term are necessary for the skyrmions to reasonably approximate many baryon features, we will use throughout this paper the Lagrangian density given by $\Lagr = \Lagr_\Sigma + \Lagr_{Skyrme}$.


\section{The Littlest  Skyrmion}

\subsection{Gauge invariant topological charge}

The expression for the winding number \eqref{equ:windingnumber} is obviously invariant under the global $SU(5)$ symmetry, however it is not invariant under the local $[SU(2)\times U(1)]^2$ gauge symmetry of the littlest Higgs model. It is thus desirable to find an expression similar to \eqref{equ:windingnumber} which is gauge invariant. An additional complication arises because we can not directly compute the winding number for a field configuration $\Sigma \in SU(5)/SO(5)$ but we need to take the detour using a field $\Phi \in SU(5)$. 
However the representation of $\Sigma$ in terms of $\Phi$ is not unique. In particular, the matrix $\Phi R$ where $R$ belongs to $SO(5)$ yields the same $\Sigma$. So the topological charge $B$ in the coset has to be an integral containing $\Phi$, $A_\mu$ and derivatives thereof, and must satisfy the following conditions:
\begin{enumerate}[(i)]
\item
invariance under the global $SU(5)$ symmetry $\Phi \to L \Phi$, such that $\Sigma \to L \Sigma L^T$ transforms as required,
\item
invariance under gauge transformations
\begin{equation*}
	\Phi \to V(x) \Phi\,, \qquad A_\mu \to V(x) A_\mu V^\dag(x) + i V(x) \partial_\mu V^\dag(x)\,,
\end{equation*}
where $V(x)$ belongs to the $[SU(2) \times U(1)]^2$ gauge group,
\item
invariance under a local $SO(5)$ symmetry $\Phi \to \Phi R(x)$, with $R(x) R^T(x) = \identity$,
\item
in the limit of vanishing gauge fields, one recovers the winding number \eqref{equ:windingnumber},
\item
time-conservation $\partial_0 B = 0$.
\end{enumerate}

Gauge invariance can be made explicit by introducing a covariant derivative for $\Phi$: $D_\mu \Phi \equiv \partial_\mu \Phi - i A_\mu \Phi$. With this notation, we have $D_\mu \Sigma = (D_\mu \Phi) \Phi^T + \Phi (D_\mu \Phi)^T$. However, it is not sufficient to promote the normal derivatives of eq.~\eqref{equ:windingnumber} to covariant derivatives in order to obtain the correct topological charge.

Instead, let's consider the current\footnote{The gauge invariant topological current is built in a similar fashion as in the $U(1)$ gauged Skyrme model of ref.~\cite{Piette:1997ny}.}
\begin{eqnarray}
	B^\mu & = & \frac{1}{24 \pi^2} \epsilon^{\mu\nu\rho\sigma} \left[ -\tr (\Phi^\dag D_\nu \Phi) (\Phi^\dag D_\rho \Phi)
		(\Phi^\dag D_\sigma \Phi)
		+ \frac{3}{2} i \tr F_{\nu\rho} \Phi D_\sigma \Phi^\dag \right] \\
		& = & \frac{1}{24 \pi^2} \epsilon^{\mu\nu\rho\sigma} \left[ -\tr (\Phi^\dag \partial_\nu \Phi) (\Phi^\dag \partial_\rho \Phi)
			(\Phi^\dag \partial_\sigma \Phi) + 3 i \tr \partial_\nu ( A_\rho \Phi \partial_\sigma \Phi^\dag ) \right. \nonumber \\
		&& \left. \quad\quad\quad\quad\quad -3 \tr (\partial_\nu A_\rho) A_\sigma + 2i \tr A_\nu A_\rho A_\sigma \right], \nonumber
	\label{equ:topologicalcurrent}
\end{eqnarray}
where $F_{\mu\nu} = \partial_\mu A_\nu - \partial_\nu A_\mu + i [A_\mu,A_\nu]$.
In its first form, this current is obviously gauge invariant and symmetric under global $SU(5)$ transformations. Defining
\begin{equation}
	B = \int d^3x B^0 = \eta(\Phi) - \frac{1}{8\pi^2} \epsilon_{ijk} \int d^3x
		\tr \left[ (\partial_i A_j) A_k - \frac{2}{3} i A_i A_j A_k \right],
	\label{equ:topologicalcharge}
\end{equation}
where we have eliminated surface terms, we see that we recover the winding number when the gauge fields are set to zero. The last term is also known as the Chern-Simons three-form.  Under a local $SO(5)$ transformation $\Phi \to \Phi R(x)$, we have
\begin{equation}
	B \to B + \eta(R) = B + 2 k, \quad k \in \mathbb Z,
\end{equation}
so that $(B \mod 2)$ satisfies the conditions (i) to (iv) above. However, this quantity is not conserved in time. From eq.~\eqref{equ:topologicalcurrent}, one has that 
\begin{equation}
	\partial_\mu B^\mu = -\frac{1}{8 \pi^2} \epsilon^{\mu\nu\rho\sigma} \tr\left[ (\partial_\mu A_\nu) (\partial_\rho A_\sigma)
		-2i (\partial_\mu A_\nu) A_\rho A_\sigma \right]
		= -\frac{1}{16 \pi^2} \tr F_{\mu\nu} \tilde F^{\mu\nu},
\end{equation}
where $\tilde F^{\mu\nu} = \frac{1}{2} \epsilon^{\mu\nu\rho\sigma} F_{\rho\sigma}$ is the dual field strength, and thus
\begin{equation}
	\partial_0 B = -\frac{1}{16 \pi^2} \int d^3 x \tr F_{\mu\nu} \tilde F^{\mu\nu}.
\end{equation}

The integrand is nevertheless a total derivative: $\tr F_{\mu\nu} \tilde F^{\mu\nu} = \tr \partial_\mu K^\mu$, with $K^\mu = 2 \epsilon^{\mu\nu\rho\sigma} [ (\partial_\nu A_\rho) A_\sigma - \frac{2}{3} i A_\nu A_\rho A_\sigma ]$. Defining $N = \frac{1}{16 \pi^2} \int d^3x \tr K^0$, the quantity $B+N$ is conserved in time. $N$ is not gauge-invariant, but $\partial_0 N$ is, and in particular $\int dt ~ \partial_0 N = N|_{t=\infty} - N|_{t=-\infty} = \frac{1}{16 \pi^2} \int d^3 x \tr F_{\mu\nu} \tilde F^{\mu\nu}$ is an integer number counting the number of instantons \cite{Manton:2004tk}.
In the particular case of the littlest Higgs model, we have two commuting $SU(2)$ gauge groups. Therefore, two different types of instantons might be present, and we can define two independent quantities counting the instantons of the two $SU(2)$ gauge groups,
\begin{equation}
	N(W_\mu^{(\alpha)}) = \frac{1}{16 \pi^2} \epsilon_{ijk} \int d^3 x
		\left[ g_\alpha^2 (\partial_i W_j^{(\alpha)a}) W_k^{(\alpha)a}
			+ 2 g_\alpha^3 W_i^{(\alpha)1} W_j^{(\alpha)2} W_k^{(\alpha)3} \right],
	\label{equ:N}
\end{equation}
where $\alpha = 1$ or 2. The topological charge defined in~\eqref{equ:topologicalcharge} can now be written as
\begin{equation}
	B = \eta(\phi) + N(W_\mu^{(1)}) + N(W_\mu^{(2)})\,,
\end{equation}
and satisfies (modulo 2) all the conditions (i) to (v) fixed above in the absence of instantons. 
A consequence of this is that the skyrmion is not stable but may decay through an electroweak instanton. 
At low temperatures these decays are strongly suppressed \cite{D'Hoker:1983kr,'tHooft:1976up,'tHooft:1976fv} such that the skyrmion is stable on cosmological time scales. A more precise estimate of its lifetime will be given later.\footnote{
Note that the stability of the skyrmion in $SU(N)/SO(N)$ coset models can
also be understood in terms of a microscopic theory, if the symmetry
breaking pattern arises from a technicolor-like theory with fermions in the
adjoint representation \cite{Bolognesi:2007ut,Auzzi:2008hu}.
}

\subsection{An SU(5) Skyrmion}

First we want to find the lightest field configuration of winding number one in the limit of vanishing gauge fields. We are always interested in field configurations going to the identity matrix at spatial infinity, and thus will consider maps $\Phi: S^3 \to SU(5)$ with nonzero winding number. Finding these maps is simplified by the unique fact that $S^3$ is isomorphic to $SU(2)$ - it is sufficient to find a map that winds once around a $SU(2)$ subgroup of $SU(5)$. 

One example of such a map is given by $\Phi(x) = \exp[2i F(r) \hat x_i T_i]$, where $F$ is a function of the distance $r$ to the origin, $\hat x_i = x_i / r$ are angular variables, and $T_i$, $i=1,2,3$ are the generators of a $SU(2)$ subgroup of $SU(5)$ obeying $[ T_i, T_j ] = i \epsilon_{ijk} T_k$ and $\tr T_i T_j = \frac{1}{2} \delta_{ij}$. This ansatz is the natural extension of the so-called hedgehog ansatz used in the original $SU(2)$ Skyrme model. It is said to be spherically symmetric, since it mixes the spatial indices coming from $\hat x_i$ with the group indices in $T_i$, so that a spatial rotation around the origin is equivalent to a global $SU(2)$ transformation,
\begin{equation}
	\Phi(R x) = \exp\left[ 2i F(r) R_{ij} \hat x_j T_i \right] = U \Phi(x) U^\dag,
\end{equation}
where $R_{ij} = 2 \tr T_i U T_j U^\dag$. Since the original Skyrme model is invariant under this diagonal $SU(2)$, its Lagrangian density at every point in space is automatically independent on the spatial direction and only depends on the distance $r$ from the origin, hence the so-called spherical symmetry.

The boundary condition $\Phi(r\to\infty) = \identity$ is obtained by choosing $F(r\to\infty) = 0$. For the field to be well defined at the origin, we must also have that $\Phi(x)$ does not depend on the angular variables $\hat x_i$ at $r=0$. This implies that $F(r=0) = k \pi$, $k \in \mathbb Z$. 
With this ansatz for $\Phi$ and the corresponding boundary conditions for $F$, the winding number~\eqref{equ:windingnumber} becomes
\begin{equation}
	\eta(\Phi) = -\frac{2}{\pi} \int\limits_0^\infty dr ~ \sin^2 F(r) ~ F'(r) = k.
	\label{equ:windingnumber:ansatz}
\end{equation}
Notice that the global structure of the $SU(2)$ subgroup generated by the $T_i$ enters crucially in the winding number. For a general representation of an algebra isomorphic to $\mathfrak{su}(2)$ with $\tr T_i T_j = C(R) \delta_{ij}$, where $C(R)$ is a Casimir invariant of the representation, the winding number integral~\eqref{equ:windingnumber} is $\eta = 2 C(r) k$. Hence, the elements of the algebra used in the hedgehog ansatz have to be in a representation with $C(R) = \frac{1}{2}$ in order to give a unit winding field configuration. As a counter-example, consider for instance the generators of $SO(3)$,
\begin{equation}
	R_1 = \left( \begin{array}{ccc}
		& -i & \\ i && \\ &&
	\end{array} \right),
	\quad
	R_2 = \left( \begin{array}{ccc}
		&& -i \\ && \\ i &&
	\end{array} \right),
	\quad
	R_3 = \left( \begin{array}{ccc}
		&& \\ && -i \\ & i &
	\end{array} \right),
\end{equation}
which also satisfy $[ R_i, R_j ] = i \epsilon_{ijk} R_k$ but have $\tr R_i R_j = 2 \delta_{ij}$, i.e. $C(R) = 2$.
Defining the hedgehog ansatz with generators $T_i = R_i$ gives an integer multiple of 4 for the winding number~\eqref{equ:windingnumber}. Another counter-example is the 4-dimensional antisymmetric set
\begin{equation}
	Z_1 = \frac{1}{2} \left( \begin{array}{cccc}
		& -i && \\ i &&& \\ &&& -i \\ && i &
	\end{array} \right),
	\quad
	Z_2 = \frac{1}{2} \left( \begin{array}{cccc}
		&& -i & \\ &&& i \\ i &&& \\ & -i &&
	\end{array} \right),
	\quad
	Z_3 = \frac{1}{2} \left( \begin{array}{cccc}
		&&& -i \\ && -i & \\ & i && \\ i &&&
	\end{array} \right),
\end{equation}
obeying $\tr Z_i Z_j = \delta_{ij}$, hence giving a winding number which is an integer multiple of 2. This very set of generators actually belong to the generators of $SO(N)$ for $N \geq 4$ and therefore allows us to identify $\pi_3(SU(5)/SO(5))$ as $\mathbb Z_2$: the winding number of any map $S^3\to SU(5)/SO(5)$ can be raised or lowered by an integer multiple of two applying the $SO(5)$ transformation $R(x) = \exp[2i F(r) \hat x_i Z_i]$, which has winding number 2 by construction.

\subsection{The SU(5)/SO(5) Skyrmion}
\label{ungauged_solution}

A generator of $\pi_3(SU(5)/SO(5))$, and thus a candidate for the skyrmion in the $SU(5)/SO(5)$ coset, is now obtained by defining
\begin{equation}
	\Sigma(x) = \Phi(x) \Phi(x)^T,
\end{equation}
where $\Phi(x)$ is an $SU(5)$ matrix of winding number one constructed using the hedgehog ansatz defined in the previous section. There is however a subtle point in using this ansatz, namely that it does not in general represent a spherically symmetric field configuration in the coset. This can be seen as follows. While the Skyrme Lagrangian for $SU(N)$ can be entirely written as a trace of products of the currents $X_\mu = \Phi^\dag \partial_\mu \Phi$, the littlest Higgs Lagrangian also contains the transpose of this current. With the above definition of $\Sigma$ we have that $\Sigma^\dag \partial_\mu \Sigma = \Phi^* [ (\Phi^\dag \partial_\mu \Phi) + (\Phi^\dag \partial_\mu \Phi)^T ] \Phi^T$, so that the Lagrangian becomes
\begin{equation}
	\Lagr = \Lagr_\Sigma + \Lagr_{Skyrme} = -\frac{f^2}{4} \tr(X_\mu+X_\mu^T)(X^\mu+X^{\mu T})
		+ \frac{1}{32 e^2} \tr \left[ X_\mu + X_\mu^T, X_\nu + X_\nu^T \right]^2 \,.
	\label{equ:LofX}
\end{equation}
Under a rotation in space, the current $X_\mu$ transforms as $X_\mu \to U X_\mu U^\dag$, where $U$ belongs to the $SU(2)$ subgroup of $SU(5)$ defined by the generators $T_i$. The Lagrangian above is obviously not rotational invariant in general. In particular taking a trivial embedding of the Pauli matrices in a $5\times5$ matrix does not preserve the spherical symmetry. 

It is well accepted that for solitons, the field configurations of highest symmetry tend to yield the lowest energy solutions to the field equations. This is the case of the original skyrmion solution \cite{Adkins:1983ya}, of the 't~Hooft-Polyakov monopole \cite{'tHooft:1974qc,Polyakov:1974ek}, and of the Julia-Zee dyon \cite{Julia:1975ff}. All these solitonic field configurations are spherically symmetric, reflecting the rotational invariance of the Lagrangian density.\footnote{Notice however that this is not the case of solutions of winding number larger than one. For monopoles, it has been proved in ref.~\cite{Weinberg:1976eq} that only the unit winding number solutions preserve the spherical symmetry, since the mass of a spherically symmetric configuration of magnetic charge $n > 1$ is larger than $n$ times the mass of the solution of magnetic charge one.} Since the above ansatz is not spherically symmetric, one may therefore wonder whether a different field configuration of higher symmetry exists, and whether it has a lower energy than the above solution. 

For the $SU(3)/SO(3)$ coset a spherically symmetric ansatz was found in \cite{Balachandran:1982ty}. In this case, an $SU(3)$ matrix $\Phi_2(x)$ is built such that a spatial rotation has the same effect on it as a $SO(3)$ transformation, namely $\Phi_2(R x) = R \Phi_2(x) R^T$, so that the field $\Sigma_2 = \Phi_2 \Phi_2^T$ transforms as $\Sigma_2(R x) = R \Sigma_2(x) R^T$ and the spherical symmetry is preserved. This ansatz was used to construct a spherically symmetric skyrmion in the littlest Higgs model in~\cite{Joseph:2009bq}. It is however unclear how to interpret this solution since the field configuration $\Phi_2$ has actually winding number two, and thus $\Sigma_2$ is a topologically trivial field configuration in the $SU(5)/SO(5)$ coset, as can be seen using eq.~\eqref{equ:winding_coset}.\footnote{Note that this is not a problem in the $SU(3)/SO(3)$ case since $\pi_3(SU(3)/SO(3)) = \mathbb Z_4$.}

\begin{figure}
	\includegraphics[width=0.48\linewidth]{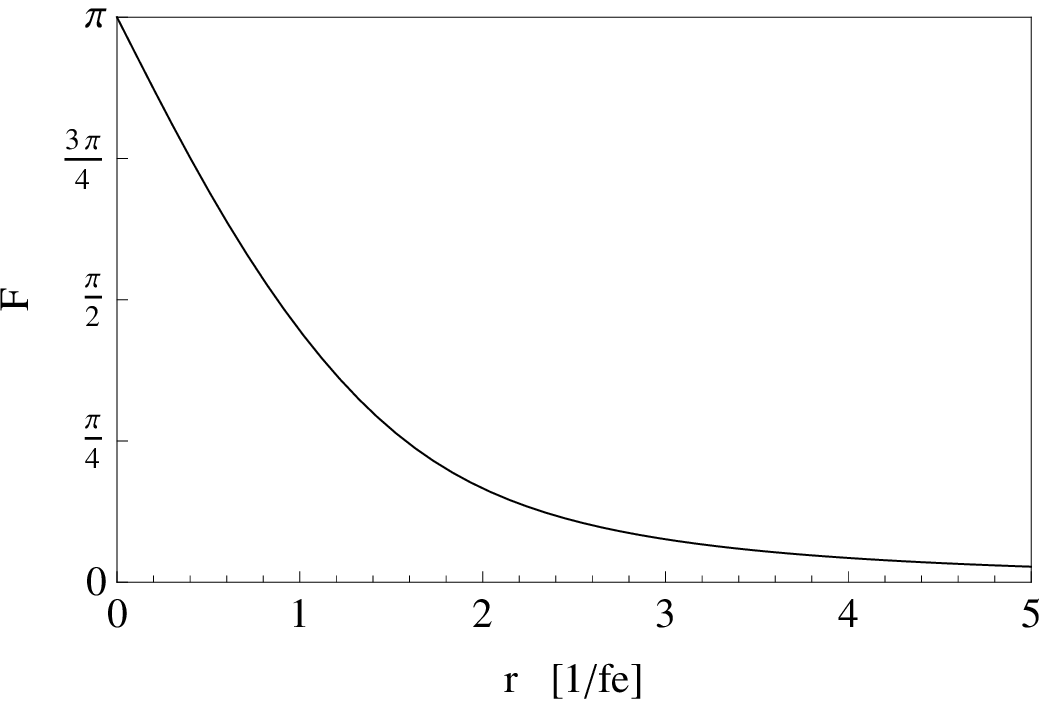}
	\includegraphics[width=0.48\linewidth]{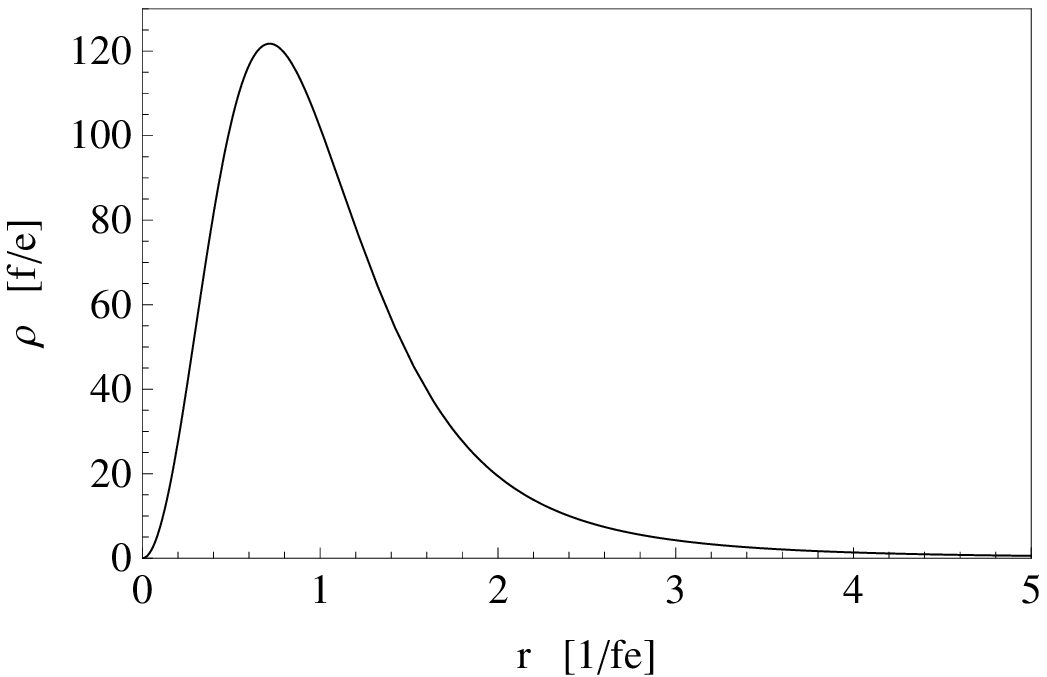}
	\caption{The profile function $F$ (left) and the radial energy density $\rho = -4\pi r^2 \Lagr$ (right) as functions of $r$.}
	\label{fig:FE_ungauged}
\end{figure}

Nevertheless, a spherically symmetric ansatz of winding number one exists for \linebreak $SU(N)/SO(N)$ with $N \geq 4$. This is because the $SU(N\geq4)$ groups are large enough to embed two commuting $SU(2)$ subgroups whose generators are transposed to each other. This is in particular realised by the gauge group of the littlest Higgs model: the $Q_i^{(1,2)}$ defined above in eq.~\eqref{equ:Q} are related through $Q_i^{(1)T}=-Q_i^{(2)}$, and satisfy $[Q_i^{(1)}, Q_j^{(2)}]=0$ and $\tr Q_i^{(1)} Q_j^{(2)} = 0$. With the ansatz
\begin{equation}
	\Phi(x) = \exp\left[ 2 i F(r) \hat x_i Q_i^{(1)} \right]
	\label{equ:ansatz}
\end{equation}
(note that taking $Q_i^{(2)}$ for the generators yields exactly the same results), the commutation relations of the group generators also imply commutations rules for the currents : $[X_\mu, X_\nu^T] = 0$. Rearranging eq.~\eqref{equ:LofX}, we obtain
\begin{equation}
	\Lagr = 2 \left( -\frac{f^2}{4} \tr X_\mu X^\mu
		+ \frac{1}{32 e^2} \tr \left[ X_\mu, X_\nu \right] \left[ X^\mu, X^\nu \right] \right).
\end{equation}
This Lagrangian is obviously invariant under spatial rotations according to the transformation rule for the currents. It is exactly twice the Lagrangian one would obtain starting from a sigma model with the $SU(5)$-valued field $\Phi(x)$ instead of $\Sigma(x)$. Therefore, the mass of the $SU(5)/SO(5)$ coset skyrmion is twice the mass of a corresponding $SU(5)$ skyrmion of winding number one. Finally, in terms of the profile function $F(r)$, the Lagrangian reads
\begin{equation}
	\Lagr_\Sigma = - f^2 \left( F'^2 + 2 \frac{\sin^2 F}{r^2} \right),
	\quad\quad
	\Lagr_{Skyrme} = -\frac{1}{e^2} \frac{\sin^2 F}{r^2} \left( 2 F'^2 + \frac{\sin^2 F}{r^2} \right),
\end{equation}
so that the energy $E = - \int d^3x ~ \Lagr$ of this static field configuration is given by
\begin{equation}
	E[F] = 4 \pi \frac{f}{e} \int\limits_0^\infty d\tilde r
		\left[ \left( \tilde r^2 + 2 \sin^2 F \right) F'^2 + \left( 2 \tilde r^2 + \sin^2 F \right) \frac{\sin^2 F}{\tilde r^2} \right].
\end{equation}
In the last equation we have performed the rescaling $r = \tilde r / (f e)$. The lowest energy configuration is then obtained by solving numerically the Euler-Lagrange equation for $F$,
\begin{equation}
	\left( \tilde r^2 + 2 \sin^2 F \right) F'' + 2 \tilde r F' + \sin 2F \left( F'^2 - 1 - \frac{\sin^2 F}{\tilde r^2} \right) = 0,
\end{equation}
with the boundary conditions $F(0)=\pi$ and $F(r\to\infty)=0$.
The profile function $F$ is shown on Fig.~\ref{fig:FE_ungauged}, together with the corresponding energy density. The mass of the ungauged skyrmion is found to be 
\begin{equation}
	M_0=145.8~f/e,
	\label{equ:mass_ungauged}
\end{equation}
twice the mass of the original $SU(2)$ skyrmion \cite{Adkins:1983ya}.\footnote{An additional factor of two is due to a difference in the normalisation between \cite{Adkins:1983ya} and the present work.}
Using eq.~\eqref{equ:topologicalcurrent} we obtain the mean square radius of the skyrmion:
\begin{equation}
	\langle r^2 \rangle \equiv \int d^3x ~ r^2 B^0 = \left( 1.058 \frac{1}{fe} \right)^2\,.
\end{equation}

The scaling of the skyrmion mass with the coefficient of the Skyrme term is particularly interesting. There is actually no upper bound on the constant $e$ from phenomenological arguments, so the mass of the skyrmion is, in principle, a free parameter of the theory. It would require knowledge of the UV completion of the littlest Higgs model to obtain an estimate of its value.
Assuming a QCD like UV completion it might be reasonable to use $e \sim 5$, which is obtained from a fit to nucleon properties \cite{Adkins:1983ya}. 
With a symmetry breaking scale $f$ around 500~GeV this gives a mass of the order of 15~TeV for the skyrmion. 

Naive dimensional analysis (NDA) gives a pre-factor of $c_s/(4\pi)^2$ for the Skyrme term where $c_s$ is an order one coefficient. This also seems to motivate $1 \lesssim e \lesssim 10$. We will see in the next chapter that the inclusion of gauge interactions modifies the dependence of the skyrmion mass on the parameter $e$.

\subsection{Gauged solution}
\label{gauged_solution}

Turning on the gauge fields can only reduce the mass of the skyrmion. We are actually looking for field configurations which have $\eta(\Phi) = 1$ and are topologically equivalent to a configuration with zero gauge fields, so that they satisfy $N(W_\mu^{(1)}) = N(W_\mu^{(2)}) = 0$. This choice will ensure a non-trivial topological charge $B = 1$. Such a configuration can be gauge-equivalent to another configuration with $\eta(\Phi) = 0$ and $N = 1$, but it would take a huge amount of time for the first configuration to evolve into the second, so that we can consider the first case to be quasi-stable.

The condition $N = 0$ implies that one can always perform a gauge transformation $A_0 \to V A_0 V^\dag + i V \partial_0 V^\dag = 0$ with $\eta(V) = 0$, so that $\eta(\Phi)$ is unchanged. In other words, we can always work in the temporal gauge $A_0 = 0$.

The ansatz~\eqref{equ:ansatz} used for the ungauged solution above spans only a $4\times4$ block of the whole $SU(5)$. While the embedding has no influence when the gauge fields are set to zero, it does have an importance for non-vanishing gauge fields. If the generators of the $SU(2)$ subgroup used in the ansatz~\eqref{equ:ansatz} do not match the gauge generators~\eqref{equ:Q}, the spherical symmetry would be broken by the gauge fields. We therefore assume that the lowest energy configuration is indeed correctly described by the ansatz~\eqref{equ:ansatz} made above.

Since the $U(1)$ gauge generators $Y^{(1,2)}$ commute with the $SU(2)$ gauge generators, the contribution of the $B_\mu$ and $\bar B_\mu$ fields to the field energy is simply given by their mass term. In order to reach the lowest energy configuration, $\bar B_\mu$ has then to be zero everywhere, while $B_\mu$ being massless is free and does not contribute to the mass of the skyrmion. An ansatz for the fields $W_\mu^a$ and $\bar W_\mu^a$ preserving the spherical symmetry can be made by writing the most general tensor decomposition in terms of the angular variables $\hat x_i$:
\begin{eqnarray}
	W_i^a & = & \frac{1}{g r} \left [ \left( \delta_{ia} - \hat x_i \hat x_a \right) a_1(r)
		+ \hat x_i \hat x_a a_2(r) + \epsilon_{iak} \hat x_k a_3(r) \right], \nonumber \\
	\bar W_i^a & = & \frac{1}{g r} \left [ \left( \delta_{ia} - \hat x_i \hat x_a \right) b_1(r)
		+ \hat x_i \hat x_a b_2(r) + \epsilon_{iak} \hat x_k b_3(r) \right].
	\label{equ:gaugeansatz}
\end{eqnarray}
Note that the factor $1 / g r$ is purely conventional, and that we work in the temporal gauge, so $W_0^a = \bar W_0^a = 0$. Plugging this ansatz into the Lagrangian, we obtain
\begin{equation}
	\Lagr_\Sigma = -f^2 \frac{1}{r^2} \left( 2 A + B \right),
	\quad\quad
	\Lagr_{Skyrme} = -\frac{1}{e^2} \frac{1}{r^4} A \left( A + 2 B \right),
\end{equation}
where we have defined
\begin{equation}
	A= \left[ (1 + a_3) \sin F - b_1 \cos F \right]^2
		+ \left[ a_1 \sin F + b_3 \cos F \right]^2,
	\quad\quad
	B = \left[ r F' - b_2 \right]^2.
\end{equation}
The spherical symmetry is here completely explicit, since the Lagrangian density depends only on $r$. The Lagrangian also contains the usual kinetic term for the gauge fields,
\begin{equation}
	\Lagr_{YM} = -\frac{1}{4} F_{ij}^a F_{ij}^a - \frac{1}{4} \bar F_{ij}^a \bar F_{ij}^a
\end{equation}
where
\begin{eqnarray}
	F_{ij}^a & = & \partial_i W_j^a - \partial_j W_i^a + g \epsilon^{abc} (W_i^b W_j^c + \bar W_i^b \bar W_j^c), \\
	\bar F_{ij}^a & = & \partial_i \bar W_j^a - \partial_j \bar W_i^a + g \epsilon^{abc} (W_i^b \bar W_j^c + \bar W_i^b W_j^c).
\end{eqnarray}
With our ansatz, this is
\begin{eqnarray}
	\Lagr_{YM} & = & - \frac{1}{g^2 r^4} \left[ \left( r a_1' - a_2 (1+a_3) - b_2 b_3 \right)^2
		+ \left( r a_3' + a_1 a_2 + b_1 b_2 \right)^2 \right. \nonumber \\
	&& \hspace{1.5cm} + \left. \left( r b_1' - b_2 (1+a_3) - a_2 b_3 \right)^2 + \left( r b_3' + a_1 b_2 + b_1 a_2 \right)^2 \right. \nonumber \\
	&& \hspace{1.5cm} + \left. \frac{1}{2} \left( a_1^2 + (1+a_3)^2 + b_1^2 + b_3^2 - 1 \right)^2
		+ 2 \left( a_1 b_1 + (1+a_3) b_3 \right)^2 \right].
\end{eqnarray}

The lowest energy configuration is then obtained by solving the corresponding Euler-Lagrange equations. One should not forget however that the winding numbers $N$ for the gauge fields as defined in eq.~\eqref{equ:N} have to remain zero. This translates into the following constraints on the profile functions:
\begin{eqnarray}
	\int\limits_0^\infty dr
		\left[ a_1 a_3' - a_1' a_3 + \frac{a_2}{r} \left( a_1^2 + (1+a_3)^2 + b_1^2 + b_3^2 - 1 \right) \right. && \nonumber \\
	\left. + b_1 b_3' - b_1' b_3 + 2 \frac{b_2}{r} \left( a_1 b_1 + (1 + a_3) b_3 \right) \right] & = & 0,
	\label{equ:constraint:1} \\
	\int\limits_0^\infty dr
		\left[ a_1 b_3' - a_1' b_3 + \frac{b_2}{r} \left( a_1^2 + (1+a_3)^2 + b_1^2 + b_3^2 - 1 \right) \right. && \nonumber \\
	\left. + b_1 a_3' - b_1' a_3 + 2 \frac{a_2}{r} \left( a_1 b_1 + (1 + a_3) b_3 \right) \right] & = & 0.
	\label{equ:constraint:2}
\end{eqnarray}
The profile functions $a_i(r), b_i(r)$ are moreover constrained by the form of the ansatz~\eqref{equ:gaugeansatz}. To obtain a finite energy solution, $a_i(r), b_i(r)$ must approach a constant value as $r\to \infty$. Definiteness at the origin furthermore implies $a_3(0) = 0$ and $a_1(0) = a_2(0)$, and similarly for the $b_i(r)$.

The Euler-Lagrange equations for $a_1$, $a_2$ and $b_3$ are satisfied by setting these three fields to zero. With this choice, the constraint~\eqref{equ:constraint:1} is automatically fulfilled. There is then a non-trivial solution with zero-energy corresponding to $a_3 = -(1 + \cos F)$, $b_1 = -\sin F$ and $b_2 = r F'$, but this solution does not satisfy eq.~\eqref{equ:constraint:2}, the left-hand side being non-zero. There are actually two obvious ways to satisfy the constraint~\eqref{equ:constraint:2}:
\begin{enumerate}[(I)]

\item
The first possibility it to set $a_3 = b_1 = 0$, and turn on only $b_2$. In this case the energy functional becomes
\begin{equation}
	E_I[F,b_2] = 4 \pi \frac{f}{e} \int\limits_0^\infty d\tilde r \left[
	\sin^2 F \left(2+ \frac{\sin^2 F}{\tilde r^2} + \frac{2 (\tilde r F' - b_2)^2}{\tilde r^2} \right)
	+ (\tilde r F' - b_2)^2 + \frac{e^2}{g^2} \frac{b_2^2}{\tilde r^2} \right],
	\label{equ:E_gauged}
\end{equation}
where we have used again the rescaled variable $\tilde r = f e \, r$. Since the energy functional does not depend on the derivative of $b_2$, the Euler-Lagrange equation for $b_2$ yields directly
\begin{equation}
	b_2(r) = \tilde r F' \left( 1 - \frac{1}{1 + (g/e)^2 (\tilde r^2 + 2  \sin^2 F)}
\right).
	\label{equ:b2}
\end{equation}
$b_2$ automatically satisfies its boundary conditions. Substituting into eq.~\eqref{equ:E_gauged}, we get
\begin{equation}
	E_I[F] = 4 \pi \frac{f}{e} \int\limits_0^\infty d\tilde r \left[ \sin^2 F \left( 2 + \frac{\sin^2 F}{\tilde r^2} \right)
		+ F'^2 \frac{(\tilde r^2 + 2 \sin^2 F)}{1 + (g/e)^2 (\tilde r^2 + 2 \sin^2 F)} \right],
	\label{equ:ansatz:I}
\end{equation}
and the Euler-Lagrange equation for $F$ becomes
\begin{eqnarray}
	\left( 1 + (g/e)^2(\tilde r^2 + 2 \sin^2 F) \right) \left( \tilde r^2 + 2 \sin^2 F \right) F'' + 2 \tilde r F' \quad && \nonumber \\
		+ \sin2F \left[ F'^2 - \left( 1 + (g/e)^2 (\tilde r^2 + 2 \sin^2 F) \right)^2 \left( 1 + \frac{\sin^2 F}{\tilde r^2} \right) \right]
		& = & 0.
	\label{equ:Euler-Lagrange:I}
\end{eqnarray}
As one can see, the mass of the skyrmion scales with $f$ and $1/e$, but depends also in a non-trivial way on the ratio $g/e$. The gauge coupling $g$ is fixed by its standard model value. For the numerical studies we use $g = 0.653$.\footnote{We use the value of $g$ at the scale $\mu = M_Z$. Our results do not depend significantly on this choice of scale, in particular in the region of interest corresponding to $e \gtrsim 1$.}
The solution of the Euler-Lagrange equation for $F(r)$ with $F(0)=\pi$ and $F(r\to\infty)=0$ is shown on Fig.~\ref{fig:F_gauged} for different values of $e$. Notice in this case that at large values of $r$, $F$ is now vanishing exponentially as $F(r) \propto \exp(-\sqrt 2 f g r)$, in strong contrast with the ungauged solution where $F$ decreases as $1/r^2$.
\begin{figure}
	\includegraphics[width=0.48\linewidth]{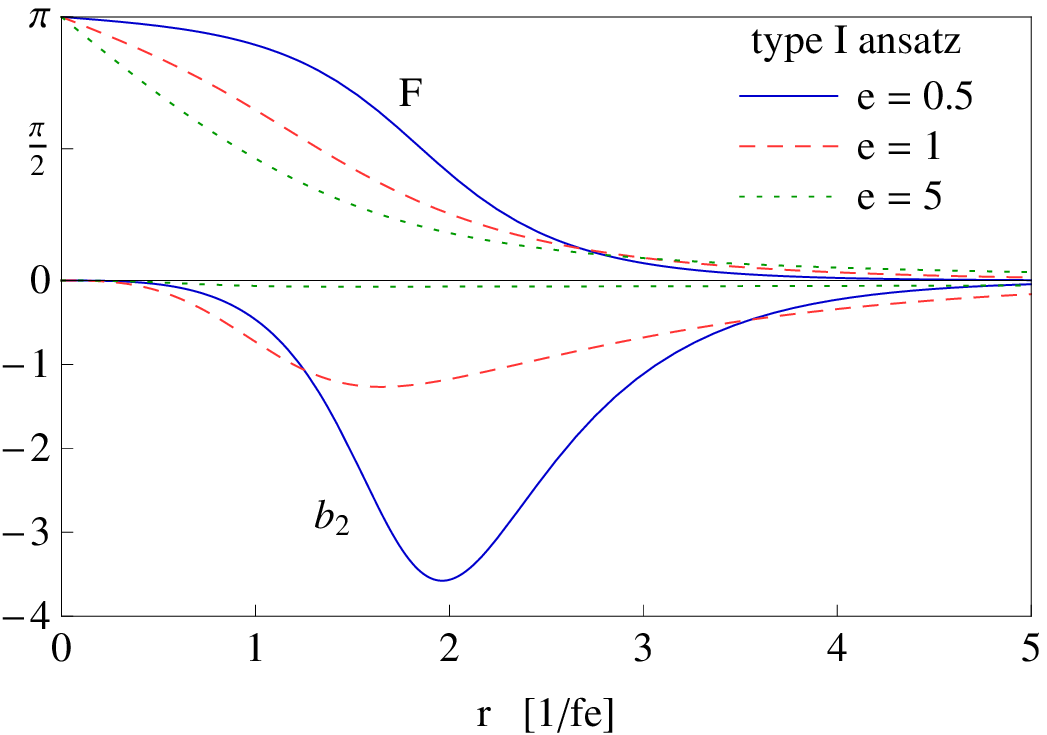}
	\includegraphics[width=0.48\linewidth]{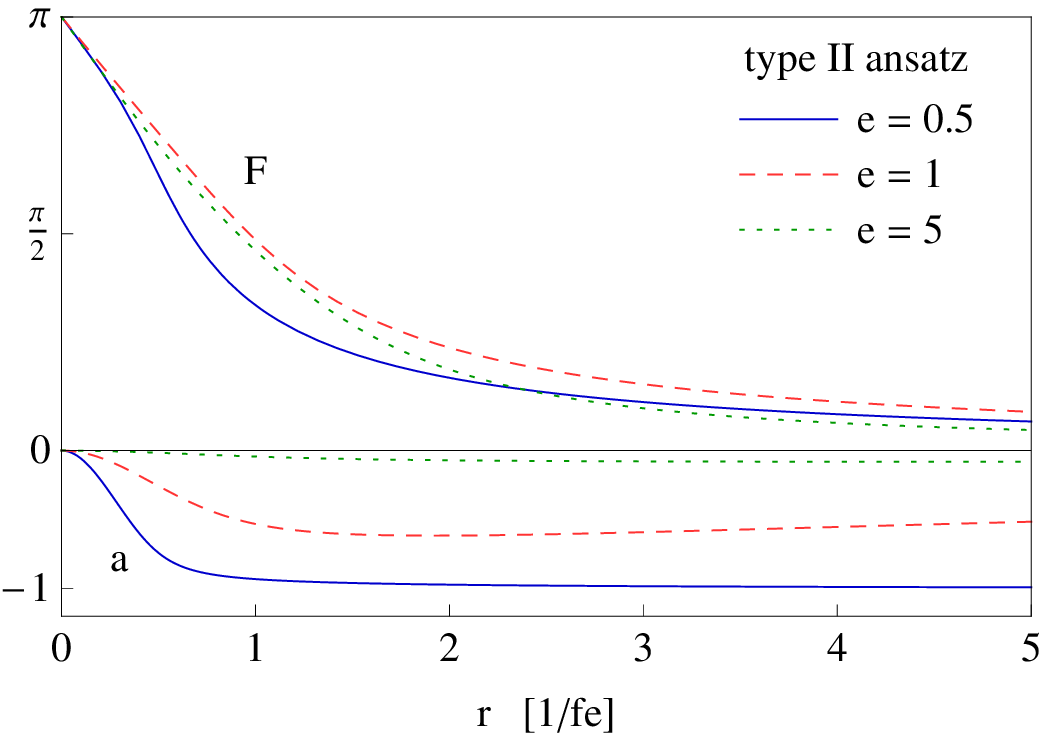}
	\caption{The profile functions $F(r)$, $b_2(r)$ and $a(r)$ for different values of the parameter $e$, for ansatz I (left) and ansatz II (right). On the right-hand side, the parameter $\omega$ is chosen to yield the lowest possible mass, namely $\omega = 0$ for $e = 0.5$, $\omega = -0.27$ for $e = 1$ and $\omega = -1.13$ for $e = 5$.}
	\label{fig:F_gauged}
\end{figure}

\item
The alternative consists in setting $b_2 = 0$ and fixing $a_3$ and $b_1$ to be proportional to each other: $a_3(r) = a(r) \cos\omega$, $b_1(r) = a(r) \sin\omega$, where $\omega$ is an arbitrary constant parameter. The energy functional is then
\begin{equation}
	E_{II}[F,a] = 4 \pi \frac{f}{e} \int\limits_0^\infty d\tilde r \left[ 
		\left( \tilde r^2 + 2 C^2 \right) F'^2 + 2 C^2 + \frac{C^4}{\tilde r^2}
		+ \frac{e^2}{g^2} \left( a'^2 + \frac{a^2 (a + 2\cos\omega)^2}{2 \tilde r^2} \right)\right],
	\label{equ:ansatz:II}
\end{equation}
where we denoted $C = \sin F + a \sin (F-\omega)$. The corresponding Euler-Lagrange equations for $F(r)$ and $a(r)$ are
{\setlength\arraycolsep{1.4pt}
\begin{eqnarray}
	\left( \tilde r^2 + 2 C^2 \right) F '' + 2 \tilde r F' + 4 \sin(F-\omega) C F' a' \hspace{4.8cm}
		&& \nonumber \\
		+ 2 C (\cos F + a \cos(F-\omega)) \left( F'^2 - 1 - \frac{C^2}{\tilde r^2} \right) & = & 0,
		\quad\quad\quad \\
	a'' - \frac{a (a + \cos\omega) (a + 2 \cos\omega)}{\tilde r^2}
	- \left(\frac{g}{e}\right)^2 2 C \sin(F-\omega) \left( F'^2 + 1 + \frac{C^2}{\tilde r^2} \right)
		& = & 0. \quad\quad\quad
\end{eqnarray}}

The numerical solutions for $F(r)$ and $a(r)$ with $F(0)=\pi$, $F(r\to\infty)=0$, $a(0)=0$ and $a(r\to\infty)=const.$ are shown in Fig.~\ref{fig:F_gauged}. In general, the dependence on $\omega$ is completely non-trivial. Nevertheless, for small $e$, the lowest mass is obtained for $\omega \cong 0$ and the profile function $a(r)$ goes to -1 as $r$ goes to infinity.
\begin{figure}
	\includegraphics[width=0.48\linewidth]{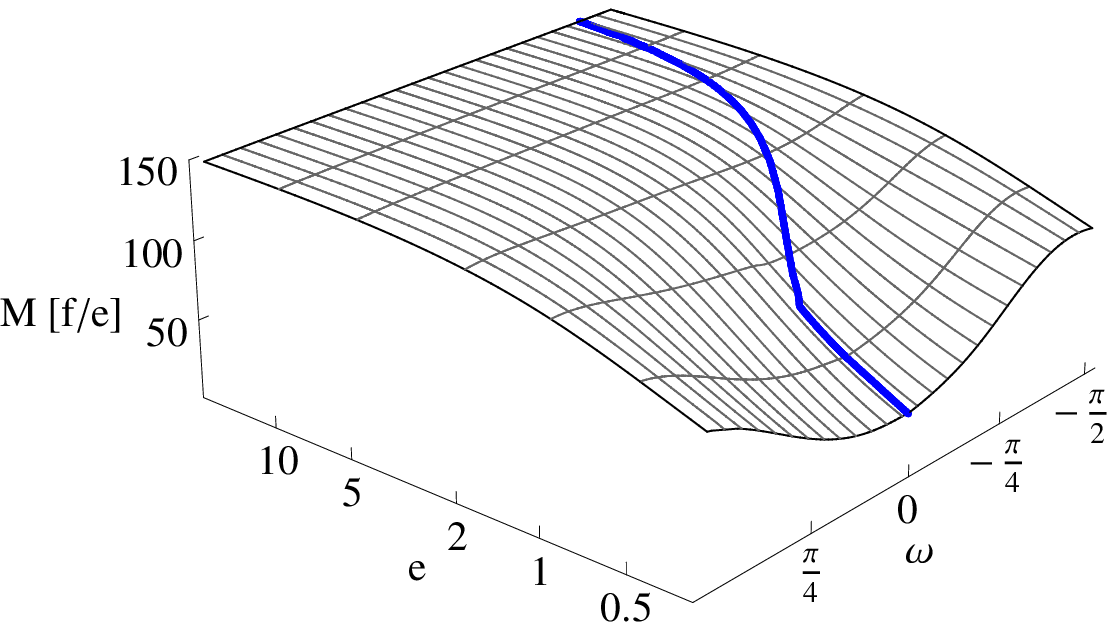}
	\includegraphics[width=0.48\linewidth]{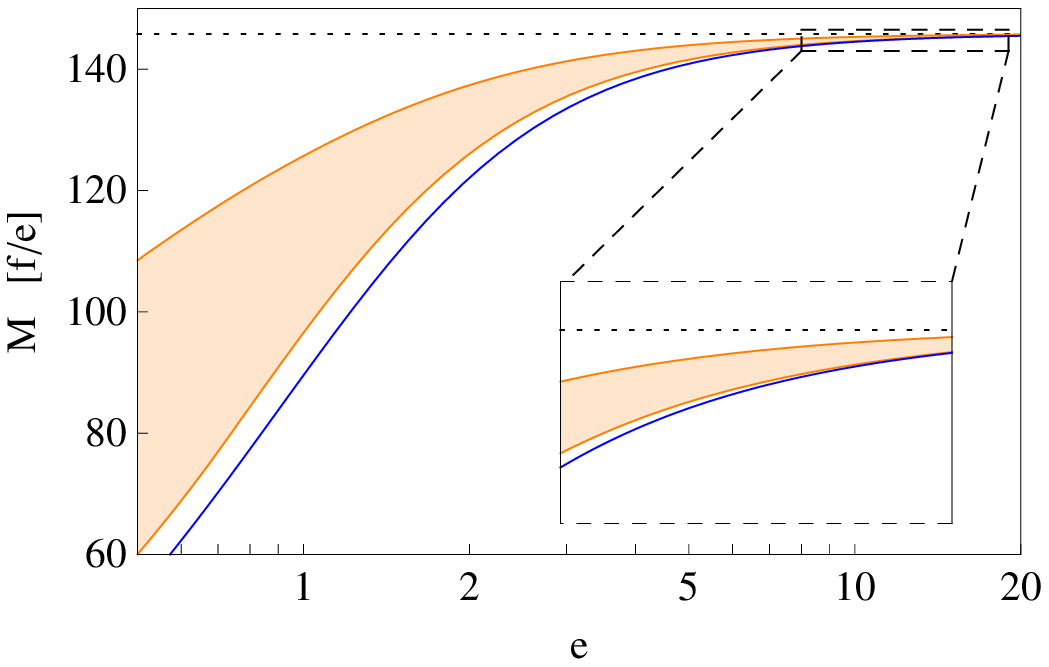}
	\caption{Left: the mass of the type II solution as a function of $e$ and $\omega$ (the thick blue line corresponds to the lowest mass for each value of $e$). Right: comparison of the ungauged solution $M_0 = 145.8 f/e$ (dotted line), the type I ansatz (blue solid line) and the type II ansatz (orange band, with $\omega$ free to vary) as functions of $e$.}
	\label{fig:M_gauged}
\end{figure}

If one chooses $\omega = 0$, only $a_3$ is turned on and our ansatz resembles the so-called Skyrme-Wu-Yang ansatz used for a $SU(2)$ gauged skyrmion in ref.~\cite{Brihaye:2004pz}.

\end{enumerate}

Although looking very different, the two ans\"atze yield very similar masses, as can be seen on Fig.~\ref{fig:M_gauged}. For $e \lesssim 10$, the lowest energy solution is obtained using the type I ansatz, while for $e \gtrsim 10$ the two choices give approximately equal masses, both very close to the ungauged case. For $e \gtrsim 5$, the mass of the gauged solution is at least 97\% of the mass of the ungauged one, the profile function $F$ is very close to the ungauged case value, and the gauge field is extremely small. In this regime, the ungauged solution can be considered a reasonable approximation.

\begin{figure}
	\includegraphics[width=0.48\linewidth]{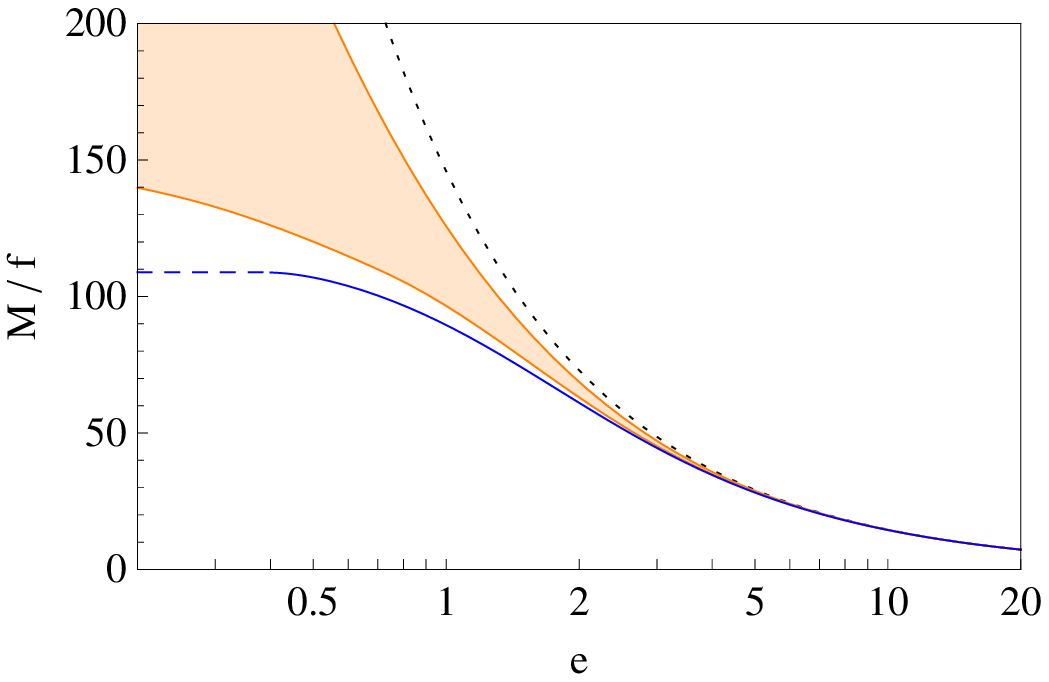}
	\includegraphics[width=0.48\linewidth]{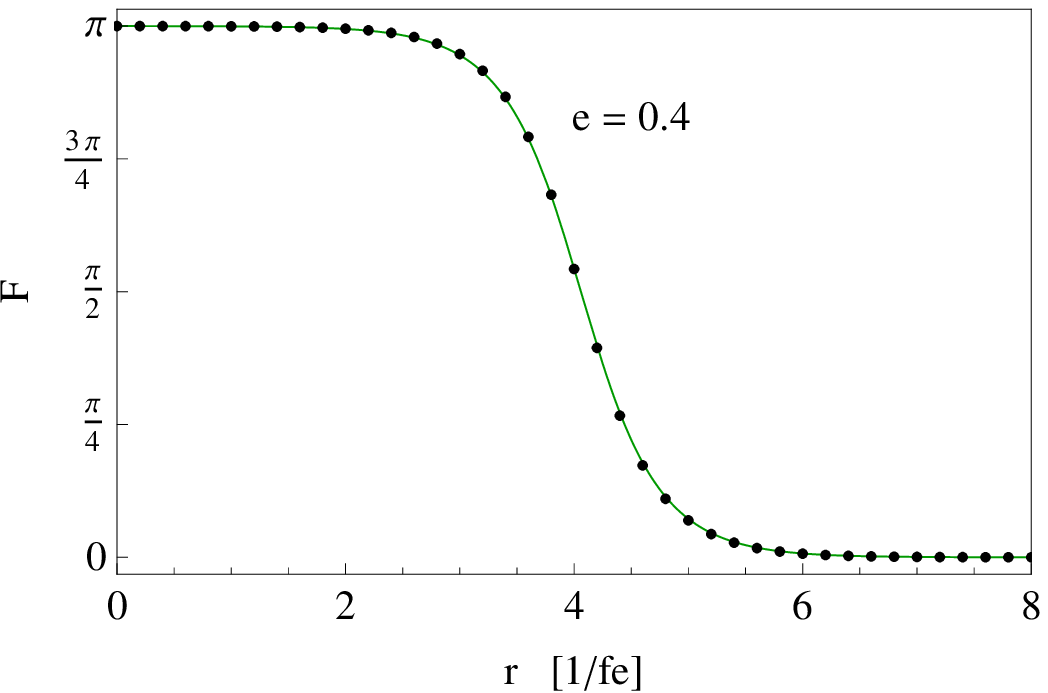}
	\caption{Left: the skyrmion mass corresponding to the ans\"atze of type I (blue solid line) and II (orange band) compared to the ungauged mass (dotted line) diverging as $1/e$; for $e \lesssim 0.5$, the convergence of the numerical method is poor, but it agrees with the analytical limit of eq.~(\ref{equ:mass_small_e}). Right: the profile function $F$ as computed numerically (black points) in the type I ansatz for $e = 0.4$, compared with the analytical result of eq.~(\ref{equ:F_small_e}) with $r_* = 4.06$ (green line).}
	\label{fig:small_e}
\end{figure}

However, the mass of the gauged skyrmion can be significantly reduced compared to the ungauged solution at small $e$. In particular, the mass of the skyrmion within the ansatz of type~I has a well defined limit at $e \to 0$, namely 
\begin{equation}
	M_{e \to 0} = 16 \sqrt 2 \pi \frac{f}{g} \cong 108.9~f,
	\label{equ:mass_small_e}
\end{equation}
as illustrated on Fig.~\ref{fig:small_e}. This bound is due to the fact that the dominance of the Skyrme term induced by a small $e$ allows for a very sharp profile for $F$, going eventually towards a step function, as explained in the appendix. In this regime, the radius of the skyrmion becomes infinitely large, but its mass remains finite. It is however still large compared to the symmetry breaking scale $f$, so that the small $e$ limit can not provide a physically interesting dark matter candidate. 

\subsection{Lifetime of the Littlest Skyrmion}

We have seen above that a gauge transformation acts on $\Phi$ as a left multiplication with a matrix $V$, where $V$ lies in the $5 \times 5$ representation of the gauge group $[SU(2) \times U(1)]^2$. Since the winding number of eq.~\eqref{equ:windingnumber} is additive, the effect of such a gauge transformation on the winding number of the skyrmion is to add a quantity equal to $\eta(V)$, the winding number of $V$. A gauge transformation with $\eta(V) = -1$ (or any odd integer value) therefore takes our skyrmion configuration to a topologically trivial field configuration.

With the ansatz~\eqref{equ:ansatz}, there is an obvious $SU(2)$ gauge transformation $V = \Phi^\dag = \exp[-2i F(r) \hat x_i Q_i^{(1)}]$ taking the winding number one field configuration $\Phi$ to the vacuum, i.e.\ the identity matrix. This field configuration takes the gauge field $W_\mu^{(1)}$ to a pure gauge configuration, $W_\mu^{(1)} = \frac{i}{g} V \partial_\mu V^\dag = \frac{i}{g} \Phi^\dag \partial_\mu \Phi$. The energy is left unchanged. The winding of $\Phi$ in $SU(5)$ is actually transferred to the gauge field: $\eta$ is taken from 1 to 0, whereas $N(W_\mu^{(1)})$ is taken from 0 to 1, so that $B$ is preserved as required.

Let us now consider a single skyrmion configuration at $t = -\infty$, namely a configuration with $\eta(\Phi) = 1$ and $N_1 = N_2 = 0$. Although we have seen that the lowest energy solution is obtained with non-zero gauge fields, we can consider here for simplicity the ungauged solution without loss of generality. As we have just seen, this configuration is equivalent to a topologically trivial scalar field configuration where the gauge field is in a pure gauge configuration, $W_\mu = \frac{i}{g} \Phi^\dag\partial_\mu \Phi$. This pure gauge configuration at $t = -\infty$ can tunnel through an instanton into a configuration with zero gauge fields at $t = +\infty$. In other words, the skyrmion can unwind with the help of an instanton. This mechanism was already studied in ref.~\cite{D'Hoker:1983kr} in the case of the standard model $SU(2)$ gauge group. The decay matrix element of the skyrmion is associated with the tunnelling probability, which was computed in \cite{'tHooft:1976up} and found to be of the order of $\exp(-8 \pi^2/g^2)$. Using this result the lifetime of the skyrmion can be estimated as\footnote{Note that the decay rate may be enhanced for large skyrmion masses, as pointed out in~\cite{Rubakov:1985it}. We do not consider those effects here, since in the concerned region of small $e$ the skyrmion mass is largely reduced due to the presence of gauge fields.}
\begin{align}
	\tau = \frac{1}{\Gamma} \sim \frac{e^{ 16\pi^2/g^2}}{M_0} \gg \tau_{\text{universe}}\,.
\end{align}

It follows that the skyrmion can be considered stable on cosmological timescales, and thus if its mass and couplings are appropriated, it can serve as a potential dark matter candidate. 

\subsection{Quantisation}

The mass of the skyrmion so far is obtained following a classical procedure. In order to compute other physical properties of the skyrmion, like its coupling to the gauge fields, one should quantise the model. The quantisation procedure is however a tedious task and will not be presented in this work. In particular, the bosonic or fermionic nature of the skyrmion may not only depend on the low-energy effective model described here, but also on the UV completion of the Littlest Higgs model. The skyrmion mass is also affected by the quantisation procedure. While the mass of the lowest skyrmion state should remain close to the classical mass computed here --- at least if the skyrmion is a scalar ---, excited states of higher mass are also expected, as in the original Skyrme model \cite{Adkins:1983ya}. Moreover, large quantum loop corrections might supposedly lower the mass of the skyrmion, although an exact computation of them is not possible \cite{Meier:1996ng}. Those issues will be addressed in a future work. 


\section{Skyrmion interactions and constraints from cosmology}

The skyrmion is a massive stable particle with at most weak couplings to standard model particles, and thus a potential dark matter candidate. A precise analysis of its properties and direct and indirect detection constraints would require the quantisation of the skyrmion, 
as mentioned in the previous section. 

However, we can already discuss some constraints coming from the skyrmion classical properties. For example, one can check if the relic abundance of skyrmions may indeed satisfy the constraints coming from cosmology. In the early universe the littlest Higgs skyrmions will be thermally produced just like any other state in the particle spectrum. In contrast to protons, they may pair annihilate due to their topological $\mathbb{Z}_2$ quantum number. Therefore, the relic abundance of skyrmions is directly determined by their annihilation cross section.

\subsection{Skyrmion-Skyrmion potential and long range forces}

Let us first show that long range forces between widely separated skyrmions are negligible at the classical level.

In the absence of gauge fields, the  potential energy binding two skyrmions can be computed precisely as long as the distance between them is much larger than their radius~\cite{Jackson:1985bn}. In this case, we can assume that they behave locally like single skyrmions and that the overall field configuration is simply described by a multiplicative ansatz $\Sigma_{12} = \Sigma_1\Sigma_2$, where $\Sigma_1$ and $\Sigma_2$ correspond to skyrmion configurations located around two points $x_1$ and $x_2$ in space, separated by a distance $d$. If the distance $d$ is of the same size as the skyrmion radius or smaller, the presence of one skyrmion will significantly distort the second skyrmion from its hedgehog shape, prohibiting an analytical calculation of the binding force between them. However, we can assume that at close distances skyrmions attract each other: due to the $\mathbb Z_2$ topology of the $SU(5)/SO(5)$ coset, the superposition of two skyrmions yields a topologically trivial field configuration, which is favoured by energy considerations.
At distances much larger than the skyrmion radius, the potential energy between the two skyrmions can be computed employing the multiplicative ansatz for the difference in energy between the two-skyrmions field configuration and two single-skyrmion configurations: $V = E[\Sigma_{12}] - E[\Sigma_1] - E[\Sigma_2]$. In this limit, the potential has been shown in~\cite{Jackson:1985bn} to be proportional to the inverse of the distance cubed: $V \sim d^{-3}$. The form of this potential is actually only determined by the large distance behaviour of the two single skyrmion solutions, which depends in turn on the asymptotic behaviour of the profile function $F(r)$. For the ungauged solution, this function scales as $1/r^2$ at large $r$.\footnote{Note that the exact large distance potential also depends on the relative isospin orientation between the two skyrmions. The sign of the interaction depends on this relative orientation, and in particular the potential vanishes if the isospins of the two skyrmions are aligned.}

For the gauged solutions, an analogous multiplicative ansatz for the two-skyrmion state cannot be employed directly, since the gauge field also contributes to the energy. However, we expect as for the ungauged solution that the potential only depends on the asymptotic behaviour of the profile functions $F$, $a_i$ and $b_i$. For the gauged type~I solution, which is always lighter than the ungauged and type~II solutions, the profile functions vanish exponentially at large $r$:
\begin{equation}
	\begin{array}{ccl}
		F(r) & \xrightarrow{r \to \infty} & c \, e^{-\sqrt{2} f g r}, \\
		b_2(r) & \xrightarrow{r \to \infty} & -\sqrt 2 \, f \, g \, c \, r \, e^{-\sqrt{2} f g r},
	\end{array}
\end{equation}
where $c$ is a numerical factor depending of the value of the Skyrme coupling $e$. We can hence safely expect the strength of the interaction to be exponentially suppressed with the distance, and therefore no large attractive or repulsive force is present at large distances, despite the fact that any two skyrmions can annihilate into light particles.
Note finally that after quantisation, the skyrmion might be charged under the electroweak gauge group, and that a potential falling off as $\exp(-m_W d)$ due to the exchange of gauge bosons can be present and of equal importance.

\subsection{Estimate of the relic density}

It is safe to assume that the skyrmions are in thermal equilibrium in the early universe, at temperatures $T>M_0$. The relic density then depends crucially on the pair annihilation cross section. A reasonable first estimate for this quantity is the geometric cross section
\begin{align}
	\sigma_A = \pi \langle r^2 \rangle \cong \frac{\pi}{(fe)^2}\,. \label{eq:geom}
\end{align}
A comparison with proton anti-proton annihilation in the original Skyrme model shows that the geometric cross section yields at least the correct order of magnitude for this process at intermediate energies  \cite{Nakamura:2010zzi}.  To parametrize the remaining uncertainty, we let the cross section vary by an order of magnitude, i.e. we take $\sigma = 10^{\pm 1}\sigma_A$ for the numerical analyses.

In addition to the cross section also the dominant final states of the annihilation process are unknown. To circumvent this problem, and to make a numerical analysis feasible, we introduce effective couplings of the skyrmion, which we assume to be a scalar, to the degrees of freedom of the littlest Higgs. To estimate the uncertainty introduced by this procedure we consider two distinct possibilities:

\begin{enumerate}[(a)]
\item 
The first possibility we consider is to couple the skyrmion directly to the Goldstone sector using the gauge invariant effective operator
\begin{equation}
	\Lagr_\text{int} = -\frac{1}{8} G_\Sigma ~ S S \tr (D_\mu \Sigma D^\mu \Sigma^\dag)\,,
	\label{equ:ssbb}
\end{equation}
where $S$ describes the skyrmion. 
This terms yields an infinite number of interactions with an arbitrary number of external legs.  For simplicity we only consider the four particle operators that mediate the annihilation of skyrmion pairs into heavy and light gauge bosons, heavy triplets $\phi$ and into little Higgses $h$. 
All of these annihilation channels give approximately the same contribution to the cross-section, as long as the mass of the final states is small compared to the skyrmion mass $M_0$:
\begin{equation}
  \sigma_{SS\to hh} = \frac{G_\Sigma^2 M_0^2}{128 \pi f^4}\,
    \frac{\beta_h (1 + \beta_h^2)^2}{\beta_S (1 - \beta_S^2)}
    \cong \frac{G_\Sigma^2 M_0^2}{32 \pi f^4}\,\frac{1}{\beta_S (1 - \beta_S^2)} \, ,
\end{equation}
and similarly for the other scalars and vector bosons.
Here, $\beta_S$ and $\beta_h$ are the relativistic velocities of the
annihilating skyrmions and the produced Higgses, respectively,
in the center-of-mass frame.
The cross-section diverges at small and large energies, in a similar fashion as the proton-antiproton annihilation cross-section. To make connection with the estimate (\ref{eq:geom}) for the total cross section, we determine the parameter $G_\Sigma$ such that the annihilation cross section for momenta $|\mathbf p| \sim \frac{1}{\sqrt{2}}M_0$, i.e. $\beta_S = \frac{1}{\sqrt 3}$, agrees with (\ref{eq:geom}). This translates into
\begin{equation}
	G_\Sigma^2 = \frac{64 \pi^2 \langle r^2 \rangle}{3 \sqrt 3 N_b} \frac{f^4}{M_0^2} \, ,
\end{equation}
where $N_b = 14$ is the number of bosons entering eq.~(\ref{equ:ssbb}). $G_\Sigma$ is independent on $f$ and $e$, due to the scaling properties of $M_0$ and $\langle r^2 \rangle$, and hence takes the constant value $G_\Sigma \cong 0.024$. The left panel of figure~\ref{fig:relic_density} shows the region in the $f-e$ plane where the skyrmion relic density agrees with the observed value. The correct dark matter abundance is obtained for relatively large values of $e$, due to the $1/e$ scaling of the geometric cross section. For small values of $f$ this corresponds to a skyrmion mass in the low TeV range, which raises some hope that these particles can be observed at the LHC. 
The freeze-out temperature and relic density were obtained using the littlest Higgs implementation of ref.~\cite{Belyaev:2006jh} in the cosmology code micrOMEGAs \cite{Belanger:2010gh}. 
\begin{figure}
	\includegraphics[width=0.48\linewidth]{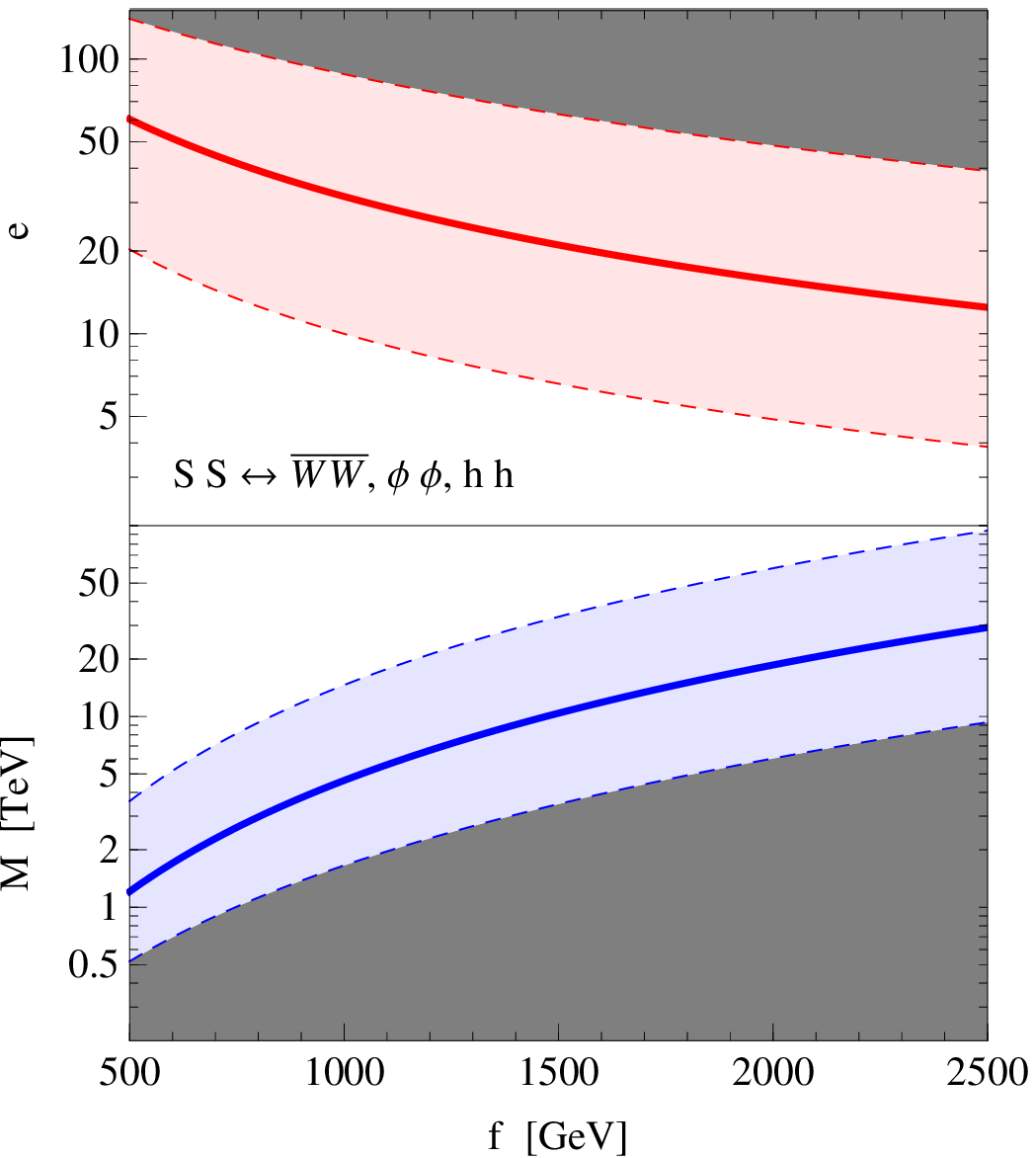}
	\includegraphics[width=0.48\linewidth]{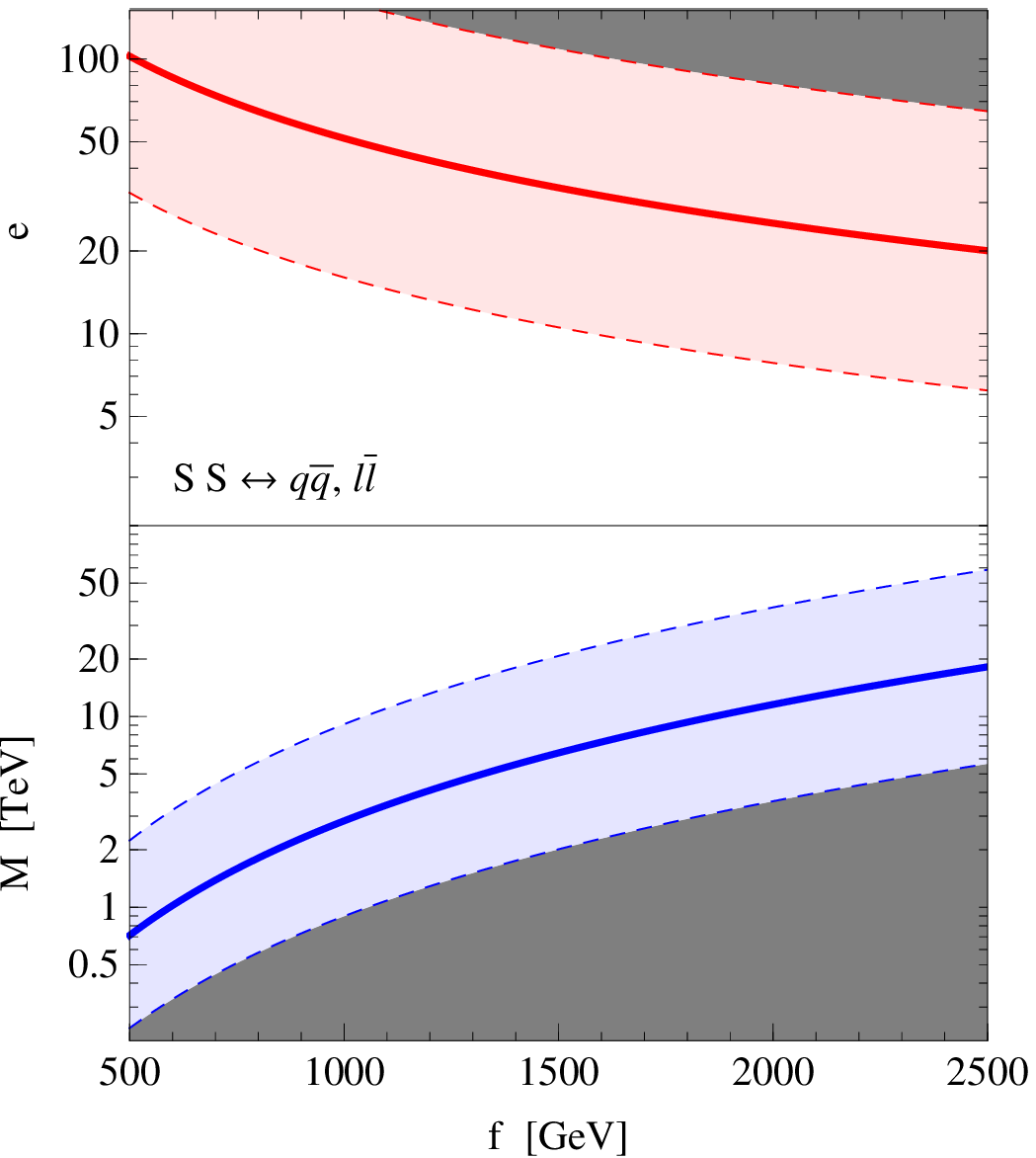}
	\caption{The value of the Skyrme coupling $e$ (upper plots) and of the corresponding skyrmion mass $M_0$ (lower plots) matching the observed dark matter relic density as a function of the symmetry breaking scale $f$. 
	The  left-hand side corresponds to coupling the skyrmion with the Nambu-Goldstone sector as in eq.~(\ref{equ:ssbb}), and the right-hand side to coupling it to the standard model quarks and leptons as in eq.~(\ref{equ:ssff}).
	The coloured band corresponds to fixing the coupling constant $G$ so that the skyrmion annihilation cross-section $\sigma$ is in the range $\frac{1}{10} \sigma_A < \sigma < 10 \sigma_A$, with the thick line corresponding to the middle value $\sigma = \sigma_A$. The dark grey regions are excluded since they predict a too large dark matter relic density.}
	\label{fig:relic_density}
\end{figure}

\item

To reduce the uncertainty from the unknown final states of the annihilation, we consider a second, purely phenomenological interaction of the form
\begin{equation}
	\Lagr_\text{int} = -\frac{1}{2} G_\psi ~ S S \bar\psi \psi\,,
	\label{equ:ssff}
\end{equation}
where again $S$ denotes the skyrmion and $\psi$ any of the standard model quarks or leptons. The coupling $G_\psi$ is taken to be
\begin{equation}
	G_\psi^2 = \frac{8 \pi^2 \langle r^2 \rangle}{N_f},
\end{equation}
where $N_f = 24$ is the number of standard model fermions. The partial cross-section into any quark or lepton pair decreases with increasing energy as
\begin{equation}
  \sigma_{SS\to\bar\psi\psi} = \frac{G_\psi^2}{8 \pi}\,\frac{\beta_\psi^3}{\beta_S}
    \cong \frac{G_\psi^2}{8 \pi}\,\frac{1}{\beta_S} \, ,
\end{equation}
where $\beta_\psi$ is the relativistic velocity of the produced fermions in the
center-of-mass frame.
The second equality holds when the mass $M_0$ of the skyrmion is much larger than the mass $m_\psi$ of the quarks and leptons.
With this choice, the sum over all standard model fermions again yields the geometric cross section. The resulting constraints on $f$ and $e$ are shown in the right panel of figure~\ref{fig:relic_density}. 
\end{enumerate}

Both models lead to similar constraints on the parameter space, the main difference being due to the different energy behaviour of the annihilation cross sections. 
Another check can be performed using the famous formula of ref.~\cite{Griest:1989wd} that relates the relic density to the annihilation cross-section of the dark matter particles,
\begin{equation}
	\Omega h^2 \cong \frac{3 \cdot 10^{-27} \text{cm}^3/\text{s}}{\langle \sigma v \rangle} \cong 0.1 .
\end{equation}
Using the naive estimate $\sigma \sim \sigma_A$,  and taking the average velocity of the skyrmions to be $v \sim \frac{1}{2} c$, this yields the constraint
\begin{equation}
	f e \sim 35~\text{TeV},
\end{equation}
which is in complete agreement with the favoured regions of Fig.~\ref{fig:relic_density}.

An important consequence of the preceding  results is that the parameter $e$ is bounded from above, which implies a lower bound on the skyrmion mass. For small values of the symmetry breaking scale $f$ the bound is rather weak, but it leads to the constraint $M_0>f$ for $f\gtrsim 1$~TeV. 
If the skyrmions were lighter than these bounds, they would be massively produced in the early universe and their annihilation cross-section would be small, so that their relic abundance at present day would exceed the observed dark matter density. Conversely, for moderate values of the Skyrme parameter, $10 \lesssim e \lesssim 100$, the skyrmion can account for the observed dark matter relic density.

There is however no lower bound on $e$ to be read from cosmological considerations. In other words, the skyrmion is allowed to be really heavy, since in this case its large mass and important annihilation cross-section makes it completely absent from our present universe. In this case, the dark matter has to be of different origin.


\section{Skyrmions in other realisations of the little Higgs}

There are a number of variations of little Higgs models in the literature which use different symmetry breaking patterns.
Depending on the third homotopy group of the coset, different types of skyrmions may emerge. For the examples to be discussed, the third homotopy groups of the various coset are computed in~\cite{Bryan:1993hz}.

\subsection*{\boldmath $SU(N)/SO(N)$}

The most prominent example of this class is the littlest Higgs model itself. 
For $N\geq 4$ we have that $\pi_3(SU(N)/SO(N)) = \mathbb{Z}_2$ and therefore models based on this coset will have skyrmion solutions very similar to those discussed in the present paper. One could in principle envision a model based on the $SU(3)/SO(3)$ coset. In that case the third homotopy group is $\mathbb{Z}_4$ and the skyrmion would be distinct from its antiskyrmion. 

\subsection*{\boldmath $SU(N)\times SU(N)/SU(N)$}

This QCD like symmetry breaking pattern with $N=3$ is realised in the minimal moose model  \cite{ArkaniHamed:2002qx,Freitas:2009jq},  where four copies of the coset are used. The third homotopy group is $\pi_3[SU(N)\times SU(N)/SU(N)] = \mathbb{Z}$. In this case skyrmion-skyrmion annihilation is not possible, and the conserved topological charge acts like the baryon number in QCD. To obtain the required relic density, it might be necessary to generate an asymmetry in the topological charge, as it is required for baryogenesis in the standard model. The advantage of such a scenario is that the dark matter density can be independent of the Skyrme parameter $e$. 

Since four copies of the coset are present in the model there will be four distinct skyrmions that carry their own conserved charge. This could have interesting consequences for dark matter searches since each skyrmion might provide a fraction of the total dark matter density in the universe.

\subsection*{\boldmath $SO(N)\times SO(N)/SO(N)$}

Examples for these models are the minimal moose with custodial symmetry \cite{Chang:2003un} with $N=5$ and the bestest little Higgs \cite{Schmaltz:2010ac} with $N=6$. For $N\geq5$ the third homotopy group is $\mathbb{Z}$. These cases are similar to the models based on $SU(N)\times SU(N)/SU(N)$.

\subsection*{\boldmath $SU(N)/Sp(N)$ and \boldmath $SU(N)/SU(N-1)$}

For these symmetry breaking patterns the third homotopy group vanishes for $N>3$. Therefore no stable skyrmions are present in these models. Note that $Sp(N)$ only exists for even $N$.


\section{Conclusions}

We have studied in detail the classical properties of skyrmions which are naturally present in most  little Higgs models. These skyrmions are the equivalent of the baryons of QCD: from the point of view of a strongly coupled UV completion, they are bound states of techni-fermions which were discussed as potential dark matter candidates already in ref.~\cite{Nussinov:1985xr}. In terms of the low-energy theory, skyrmions are built out of the numerous bosonic fields which appear in little Higgs models. 
The stability of the skyrmion is ensured by a topologically conserved quantity, a direct consequence of the various symmetry breaking patterns in little Higgs models. We showed that this quantity can be made gauge invariant, without spoiling the stability of the skyrmion on time scales larger than the age of the universe.

We found a spherically symmetric ansatz for skyrmions in $SU(N)/SO(N)$ cosets with $N\geq 4$, and solved the differential equation for the profile function. Using those results we obtained the classical mass and the radius of the skyrmion in the littlest Higgs model. Next we considered the effects of gauge interactions on the prediction of the skyrmion mass. The rich gauge structure of the littlest Higgs model leads to highly nonlinear differential equations subject to additional topological constraints. The equations are solved for two special cases, and we show that the mass of the skyrmion is reduced when gauge interactions are taken into account. We are able to establish an upper bound on the skyrmion mass in the limit of $e \to 0$, in agreement with previous findings. 

For phenomenological applications, the large $e$ regime is more interesting, since there the mass of the skyrmion can be lowered down to the 1~TeV scale. In this regime the effects of gauge interactions on the skyrmion profile can be neglected. 

We argued that long range forces are absent for pairs of skyrmions. This, together with its stability and weak scale interactions with ordinary matter, makes the skyrmion a potential dark matter candidate. Using the geometric cross section as an estimate for the annihilation rate, we determined the parameters $f$ and $e$ for which the relic density agrees with the observed dark matter density. 

The skyrmion can account for all of the dark matter in the universe, provided that $e$ is large enough. This in turn implies a relatively small mass. For low values of the symmetry breaking scale, $f \leq 1$~TeV, the skyrmion can be light enough to be produced at the LHC. 

The results obtained so far are only at the classical level. It would be of great interest to study the effects of quantisation on the skyrmion mass and on its coupling to ordinary matter. Moreover, since we saw that the skyrmion might be as light as the TeV scale, it could be produced in pairs at hadron colliders. In particular, it might be interesting to compare the results of the present work to modern models of technibaryons, both in terms of collider signatures and of consequences in cosmology.


\acknowledgments

The authors would like to thank Ayres Freitas, Cosmas Zachos and Roberto Auzzi for valuable comments on the manuscript, and Timo Schmidt and Tom Ilmanen for useful discussions. This work was supported in part by the Schweizer Nationalfonds and by the U.S. Department of Energy, Division of High Energy Physics, under Contract DE-AC02-06CH11357 and DE-FG02-84ER40173.


\appendix
\section{Appendix: Small $e$ limit of the gauged solution}
\label{appendix}

The limit of small $e$ in eq.~(\ref{equ:ansatz:I}) is not easy to take numerically. For $e \gtrsim 1$ the profile function $F$ has a finite negative slope at $r = 0$, as visible on Fig.~\ref{fig:F_gauged}. In this parameter range, the Euler-Lagrange equation~(\ref{equ:Euler-Lagrange:I}) can be solved employing for example a shooting or a relaxation method. However, at small $e$, the profile function $F$ tends to be very flat around the boundary $r = 0$. Here, our shooting implementation becomes increasingly unstable. Instead, we resort to a relaxation algorithm, which produces reliable results for the profile functions down to values of $e$ not much smaller than $0.4$. For even smaller values of $e$ it becomes increasingly difficult to obtain precise results, also with the relaxation method.

Nevertheless, the limit in which $e$ goes to zero can be taken analytically by making the following observations:
\begin{enumerate}[(a)]

\item
The fraction in the last term of eq.~(\ref{equ:ansatz:I}) tends towards $e^2/g^2$ as $e \to 0$. The resulting term $e^2/g^2 (F')^2$ can however still be large, since $F$ tends to become a step function, and hence $F'$ is large around the step. 

\item
The term $\sin^4 F/r^2$ becomes subdominant in comparison with $2 \sin^2 F$, since the sine is small everywhere except around the step of $F$, and the factor $1/r^2$ makes the contribution around the step small, due to the large value of $r$ there.
\end{enumerate}
With those two observations, the energy functional (\ref{equ:ansatz:I}) can be approximated as
\begin{equation}
	E_{I,e \to 0}[F] = 4 \pi \frac{f}{e} \int\limits_0^\infty dr \left[ 2 \sin^2 F + \frac{e^2}{g^2} F'^2 \right],
	\label{equ:E_small_e}
\end{equation}
yielding the Euler-Lagrange equation for $F$
\begin{equation}
	\frac{e^2}{g^2} F'' = \sin 2F.
\end{equation}
Since this equation is indepent on $r$, it can be integrated, giving
\begin{equation}
	\frac{e^2}{g^2} F'^2 = 2 \sin^2 F + C,
\end{equation}
where $C$ is a constant. Requiring $F'(r \to \infty) = 0$ fixes the constant to $C = 0$. This value of $C$ also implies that at $r = 0$, where $F(0) = \pi$, the derivative of $F$ vanishes, as observed numerically. Since $F$ has to be decreasing between $\pi$ and 0, the equation for $F$ becomes
\begin{equation}
	F' = -\sqrt 2 \frac{g}{e} \sin F,
	\label{equ:equ_small_e}
\end{equation}
which is solved by the function
\begin{equation}
	F(r) = 2 \arctan \left[ \exp\left( -\sqrt 2 \frac{g}{e} (r - r_*) \right) \right].
	\label{equ:F_small_e}
\end{equation}
$r_*$ is a constant fixing the position of the step, which is supposedly going to infinity at small values of $e$.
However, the energy obtained with this $F$ is independent on $r_*$: using eq.~(\ref{equ:equ_small_e}), one can rewrite the energy functional~(\ref{equ:E_small_e}) as
\begin{equation}
	E_{I,e \to 0}[F] = 8 \sqrt 2 \pi \frac{f}{g} \int\limits_0^\infty dr (-\sin F) F'
		= 8 \sqrt 2 \pi \frac{f}{g} \int\limits_\pi^0 dF \cos F = 16 \sqrt 2 \pi \frac{f}{g},
\end{equation}
which is exactly the limit given in eq.~(\ref{equ:mass_small_e}). Fig.~(\ref{fig:small_e}) shows that both the mass~(\ref{equ:mass_small_e}) and the step function given by eq.~(\ref{equ:F_small_e}) are in good agreement with the full numerical solution for $e \cong 0.4$ already.


\bibliography{Bibliography}
\bibliographystyle{JHEP}

\end{document}